\pdfoutput=1
\documentclass[11pt,oneside,british,reqno]{amsart}
\usepackage{lmodern}
\usepackage{avant}
\usepackage{courier}

\usepackage[T1]{fontenc}
\usepackage[latin9]{inputenc}
\usepackage[a4paper,left=3cm, right=3cm,top=3cm, bottom=3cm]{geometry}
\usepackage{subcaption}
\usepackage{color}
\usepackage{babel}
\usepackage{units}
\usepackage{amsbsy}
\usepackage{amstext}
\usepackage{amsbsy}
\usepackage{url}
\usepackage{amsthm}
\usepackage{amssymb}
\usepackage{graphicx}
\usepackage{setspace}
\usepackage{nicefrac}
\usepackage{float}
\usepackage{bbold}
\usepackage[titletoc,title]{appendix}
\usepackage[unicode=true,pdfusetitle,
 bookmarks=true,bookmarksnumbered=false,bookmarksopen=false,
 breaklinks=false,pdfborder={0 0 1},backref=false,colorlinks=true]
 {hyperref}
\hypersetup{
 pdfborderstyle={},pdfborderstyle={},linkcolor=blue,urlcolor=blue,citecolor=blue}

\makeatletter

\allowdisplaybreaks

\numberwithin{equation}{section}
\numberwithin{figure}{section}
\numberwithin{table}{section}

\makeatother

\begin{document} 

\markboth{M. Angelelli and E. Ciavolino and P. Pasca}{Streaming Generalized Cross Entropy}

\title{Streaming Generalized Cross Entropy}

\author{M. Angelelli$^{1,2}$, E. Ciavolino$^{1}$ and P. Pasca$^{1}$}

\date{}

\maketitle
%\begin{center}
%Department of Mathematics and Physics ``Ennio De Giorgi'', 
%\par\end{center}

\begin{center}
University of Salento$^{1}$ and sezione INFN$^{2}$, 
\par\end{center}

\begin{center}
Lecce, 73100, Italy
\par\end{center}

\begin{abstract}
We propose a new method to combine adaptive processes with a class of entropy estimators for the case of streams of data. 

Starting from a first estimation  obtained from a batch of initial data, model parameters are estimated at each step by combining the prior knowledge with the new observation (or a block of observations). 
This allows to extend the maximum entropy technique to a dynamical setting, also distinguishing between entropic contributions of the signal and the error. Furthermore, it provides a suitable approximation of standard GME problems when the exacted solutions are hard to evaluate.

We test this method by performing numerical simulations at various sample sizes and batch dimensions. Moreover, we extend this analysis exploring intermediate cases between streaming GCE and standard GCE, i.e. considering \emph{blocks} of observations of different sizes to update the estimates, and incorporating collinearity effects as well. The role of time in the balance between entropic contributions of signal and errors is further explored considering a variation of the Streaming GCE algorithm, namely, Weighted Streaming GCE. Finally, we discuss the results: in particular, we highlight the main characteristics of this method, the range of application, and future perspectives. 
\end{abstract}

\vspace*{.3in}

MSC[2010] \textit{94A17, 62Jxx, 62L12}

\section{Introduction}

\label{sec: Introduction}

Several situations in modern research and technological development
have an essential time directionality, which arises from a sequential
acquisition of the data while producing outputs and decisions at intermediate
steps.

Concrete examples of this feature include the update of control parameters
in dynamical systems (mobile phones, GPS), social networks, image
analysis, economic analysis (cryptocurrency), as well as longitudinal
studies in the social sciences or psychological change occurring over
time during the psychoterapeutic process, \textit{etc}. These data
may be characterized by 
\begin{itemize}
\item large sample size (e.g., big data); 
\item unstructured data (e.g., audio/video, natural language); 
\item the need of a continuous, streaming analysis (e.g., image analysis); 
\item the need for fast answer/output given a new input (e.g., submarine
robot control). 
\end{itemize}
These aspects show many drawbacks in concrete applications, because
standard analyses on whole dataset: 
\begin{itemize}
\item are computationally expensive, as they require the whole algorithm
to run at each new observation; 
\item hide the relationship between subsequent estimates at consecutive
times $i\rightarrow i+1$ and, hence, the information acquisition
and knowledge dynamics; 
\item may fail to provide prompt responses, and have to be adapted to work
in real time with online data. 
\end{itemize}
Therefore, in these situations, one cannot disregard the role of the
time flow. The process itself has to flexibly \emph{adapt} to the
evolution of the system.

In order to highlight the temporal structure of the data, we will
make use of the Maximum Entropy Principle (MEP), which allows to extract
the essential information available in a set of observations with
the least amount of additional assumptions. The MEP is a cornerstone
in contemporary science due to its capacity to embed either quantitative
or qualitative aspects of information within models for a variety
of applications and, hence, connect several different frameworks:
nowadays, applications of the MEP can be found in artificial intelligence
(Solomonoff's universal inductive reasoning \cite{Solomonoff1964}),
complexity theory \cite{Holzinger2014}, linguistics \cite{Berger1996},
biology \cite{Gladyshev1997} \textit{etc}.

A remarkable development of the MEP in statistical regression is given
by the Generalized Maximum Entropy (GME) approach \cite{Golan1996,Golan2007,Golan2018},
which is successfully employed to provide estimates of both parameters
and errors in regression models. The extension to multivariate analysis
has been developed in the last decade \cite{Ciavolino2009a,Ciavolino2009b}.
These techniques proved to be useful in psychological and social studies
on both perception and satisfaction \cite{Ciavolino2014a,Ciavolino2014b}, in the analysis of observational studies \cite{Amusa2019} and also in combination with Structural Equation Models (SEM) \cite{Bernardini2011,Ciavolino2014c,Ciavolino2016}.

On another hand, as the implementation of the GME method involves
a whole dataset, there is a permutation symmetry for the consistency
constraints: in this context, the implications of the temporal dimension
are lost.

The aim of the present work is to investigate the combination of the
information theory based on the GME and adaptive aspects during a
streaming flow of data. This purpose entails a natural distinction
between the parameters of the model, which represent the ``recurrent''
part of the set of observations, and errors, which represent the ``accidental''
part. The parameters, which will also be referred to as the \emph{signal},
are iteratively estimated from upcoming new data, while the errors
are necessary to make the recurrent part consistent with different
instances of the underlying model. Further insight on the above-mentioned
components is evident when observing their natural evolution over
a stream of data.

In order to explore such distinction, we consider an adaptive process
and update the probability distributions associated with both the
signal and the errors. We choose the cross entropy (or Kullback-Leibler
divergence) as cost function to be minimized at each step, under the
constraints given by the new data in the stream. The asymmetry between
the signal and the error is represented by different priors associated
to each one of them in the cross entropy. These two notions, \textemdash adaptive
evolution during a stream and entropy-based principle to combine prior
information and new data \textemdash are the roots of the Streaming
Generalized Cross Entropy (Stre-GCE) presented in this work.

The inclusion of the temporal aspect can be used to explore the dynamics
of the parameter estimation when new data come up: indeed, the temporal factor can be
used to interpolate the partial information represented by a data
subset (the batch) and the full information provided by the whole
sample: the update of the weights associated to both the already existing
and the upcoming information over time will be the focus of the extension
of the Stre-GCE algorithm called \emph{Weighted} Stre-GCE.

The paper is organized as follows: in Section \ref{sec: Preliminaries},
we provide a basic definition of our algorithm by briefly mentioning
adaptive process and the standard GCE method in linear regression.
The description of the Stre-GCE method, the regime where it is considered,
its analogies and differences with respect to the GME and GCE techniques
are discussed in Section \ref{sec: Characteristics and range of application}.
Section \ref{sec: Stre-GCE algorithm} is devoted to the formalization
of the Stre-GCE algorithm. In order to compare the performance of
Stre-GCE with respect to GME and simple GCE, we run a set of simulations
that are discussed in Section \ref{sec: Simulations}. In particular,
we focus on the comparison between Root Mean Square Errors (RMSE),
we extend Stre-GCE beyond the single-unit update to a stream of blocks
of data, and we explore multicollinearity effects. In Section \ref{sec: Weighted Stre-GCE},
we extend the basic Stre-GCE algorithm so that the relation between
the signal and the error contributions evolves during the data stream.
Results of these simulations are discussed in detail in Section \ref{sec: Discussion },
while future perspectives and applications of Stre-GCE are suggested
in Section \ref{sec: Conclusion and future perspectives}.

\section{Preliminaries}

\label{sec: Preliminaries}

The implementation of the Stre-GCE algorithm requires two main ``ingredients'':
adaptive processes and the Generalized Cross Entropy method.

\subsection{Adaptive processes}

\label{subsec: Adaptive processes}

When a model evolves over time, the role of all the agents should be
taken into account (e.g. the actions of the observer, the decisions
made consistently with the previously processed data). Inclusion of
flow of information along with a sequence of control steps in the
same framework leads to an adaptive process in which previous outputs
and new data are merged together based on a given criterion. Such
a criterion is often quantified by a cost (or reward) function, that
is, a measure of the discrepancy between the actual model and a ``more
favorable'' one, in accordance with the new data. Applications include,
but are not limited to, system and control theory \cite{Simon2006}
pattern recognition \cite{Widrow1988}, and reinforcement learning
\cite{Xu2002}.

A major example of adaptive process is the Kalman Filter (KF), a fundamental
tool in control theory \cite{Simon2006}. The standard Kalman filter
takes advantage from the linearity of the model and the Gaussian distribution
of errors at its roots. Consequently, one can consider algebraic formulations
based on Least Squares (LS) methods in order to provide exact formulas
for optimal estimation. The Recursive Least Squares (RLS) algorithm
fills the need of an efficient inclusion of new data in the model.

However, the efficacy of this method is limited by the same assumptions
mentioned before. Nonlinear models and multimodal distributions make
the implementation of the standard Kalman filter ineffective \cite{Simon2006,Zanetti2012}.
This drastically impacts on decision theory, where non-trivial
symmetries may relate different states or transitions, and the standard
Kalman filter does not faithfully represent this ambiguity. An efficient
application of the KF in relation to its optimality properties among
linear estimators only relies on the exact knowledge of covariance
matrices for errors: however, this condition is hardly fulfilled in
concrete situations.

For these reasons, nonlinear extensions of the Kalman filter have
been developed (e.g. extended KF, unscented KF, particle filters),
as well as additional numerical methods (see \cite{Daum2005}). Even
though each of these approaches returns some information on the required
probability distribution with greater generality, they are often computationally
expensive and rely on recursive schemes for approximations \cite{Daum2005,Zanetti2012},
thus being unsuitable for real-time data.

Besides statistical assumptions, KF and several other methods in linear
models depend on specific algebraic hypotheses, which reflect on the
observability of the system: these requirements are not fulfilled
in presence of multicollinearity or when numerical approximations
lead to rank reductions. Multicollinearity can be managed with other
approaches {(see e.g. \cite{Bagya2018} for an analysis of performance of regression methods under multicollinearity)}, including the GME and GCE methods that are summarized
in the following subsections.

\subsection{Generalized Cross Entropy (GCE)}

\label{subsec: Generalized Maximum Entropy}

The notion of entropy arose in early works by Boltzmann, Gibbs and
Planck in statistical mechanics, where the equilibrium state in a
macroscopic system is represented by the maximum entropy distribution
compatible with macroscopic observables \cite{LL1980,Feynman1982},
and by Shannon in the analysis of optimal communication information
theory (see \cite{Cover2006} for a review). The key aspects of the
Maximum Entropy Principle (MEP) have been intensively explored by
Khinchin \cite{Khinchin1957}, Kullback and Leibler and Jaynes, in
order to identify its characterizing features that do not rely on
the specific physical system \cite{Dewar2009} and, hence, make it
a fundamental principle for inference \cite{Jaynes2003}. A number
of distinctive features of the MEP can be also drawn in relation to
other methods, such as Maximum Likelihood, Empirical Likelihood, Bayesian
method of moments and (ordinary or weighted) Least Squares \cite{Golan1998}.

For our purposes, GME and GCE approaches for regression models play
a relevant role \cite{Golan1996,Golan2007,Golan2018}. In particular,
we focus on linear regression 
\begin{equation}
\mathbf{y}=\boldsymbol{\beta}\cdot\mathbf{x}+\boldsymbol{\varepsilon}\label{eq: matrix consistency constraint}
\end{equation}
involving a set of $n$ units 
\begin{equation}
y_{i}=\sum_{j=1}^{J}\beta_{j}\cdot x_{i,j}+\varepsilon_{i},\quad i\in\{1,\dots,n\},\label{eq: consistency constraints}
\end{equation}
Through GME, one can express both the parameters $\beta_{j}$, $j\in\{1,\dots,J\}$
(and, possibly, the intercept $\alpha$) and the errors $\varepsilon_{i}$,
$i\in\{1,\dots,n\}$, as expected values with respect to a certain
probability distribution $\left(\mathbf{p}^{\beta_{1}},\dots,\mathbf{p}^{\beta_{J}},\mathbf{p}_{1}^{\varepsilon},\dots,\mathbf{p}_{n}^{\varepsilon}\right)$
supported on a finite set. Specifically, $K$ is the number of support
points $\{z_{j,k}^{\beta}:\,k\in\{1,\dots,K\}\}$ for $\beta_{j}$
and $H$ is the number of support points $\{z_{i,h}^{\varepsilon}:\,h\in\{1,\dots,H\}\}$
for $\varepsilon_{i}$. So (\ref{eq: consistency constraints}) can
be restated as a set of $n$ constraints 
\begin{equation}
y_{i}=\sum_{j=1}^{J}\sum_{k=1}^{K}p_{j,k}^{\beta}\cdot z_{j,k}^{\beta}\cdot x_{i,j}+\sum_{h=1}^{H}p_{i,h}^{\varepsilon}\cdot z_{i,h}^{\varepsilon},\quad i\in\{1,\dots,n\}\label{eq: consistency constraints GME}
\end{equation}
with 
\begin{equation}
\sum_{k=1}^{K}p_{j,k}^{\beta}=\sum_{h=1}^{H}p_{i,h}^{\varepsilon}=1,\quad j\in\{1,\dots,J\},\,i\in\{1,\dots,n\}.\label{eq: normalizations}
\end{equation}
In matricial terms, (\ref{eq: consistency constraints GME})-(\ref{eq: normalizations})
can be expressed as 
\begin{align*}
\mathbf{y}_{n\times1}= & \mathbf{X}_{n\times J}\cdot\mathrm{diag}(\mathbf{Z}_{J\times K}^{\beta}\cdot(\mathbf{P}_{J\times K}^{\beta})^{T})_{J\times1}+\mathbf{P}_{n\times H}^{\varepsilon}\cdot\mathbf{z}_{H\times1}^{\varepsilon},\\
\mathbf{1}_{J\times1}= & \mathbf{P}_{J\times K}^{\beta}\cdot\mathbf{1}_{K\times1},\\
\mathbf{1}_{n\times1}= & \mathbf{P}_{n\times H}^{\varepsilon}\cdot\mathbf{1}_{H\times1}
\end{align*}
where the dimensions of each matrix are made explicit in the corresponding
suffix. For any natural numbers $a$ and $b$, $\mathrm{diag}(\mathbf{M}_{a\times a})$
is the $(a\times1)$-dimensional matrix (a column vector) whose entries
are the diagonal entries of $\mathbf{M}$, and $\mathbf{1}_{a\times b}$
is the $(a\times b)$-dimensional matrix with all entries equal to
$1$. The maximization of Shannon's entropy 
\begin{equation}
H\left(\mathbf{p}^{\beta_{1}},\dots,\mathbf{p}^{\beta_{J}},\mathbf{p}_{1}^{\varepsilon},\dots,\mathbf{p}_{n}^{\varepsilon}\right)=-\sum_{j=1}^{J}\sum_{k=1}^{K}p_{j,k}^{\beta}\cdot\ln p_{j,k}^{\beta}-\sum_{i=1}^{n}\sum_{h=1}^{H}p_{i,h}^{\varepsilon}\cdot\ln p_{i,h}^{\varepsilon}\label{eq: GEM entropy}
\end{equation}
subject to consistency (\ref{eq: consistency constraints GME}) and
normalization (\ref{eq: normalizations}) constraints returns a highly
accurate estimation of the parameter and the error terms. Note that
the components of the joint distribution $\left(\mathbf{p}^{\beta_{1}},\dots,\mathbf{p}^{\beta_{J}},\mathbf{p}_{1}^{\varepsilon},\dots,\mathbf{p}_{n}^{\varepsilon}\right)$
are assumed to be independent, as it can be seen from the form of
Shannon's entropy (\ref{eq: GEM entropy}). In particular, we can
write $H\left(\mathbf{p}^{\beta_{1}},\dots,\mathbf{p}^{\beta_{J}},\mathbf{p}_{1}^{\varepsilon},\dots,\mathbf{p}_{n}^{\varepsilon}\right)=H\left(\mathbf{p}^{\beta_{1}},\dots,\mathbf{p}^{\beta_{J}}\right)+H\left(\mathbf{p}_{1}^{\varepsilon},\dots,\mathbf{p}_{n}^{\varepsilon}\right)$,
where 
\begin{eqnarray*}
H\left(\mathbf{p}^{\beta_{1}},\dots,\mathbf{p}^{\beta_{J}}\right) & := & -\sum_{j=1}^{J}\sum_{k=1}^{K}p_{j,k}^{\beta}\cdot\ln p_{j,k}^{\beta},\\
H\left(\mathbf{p}_{1}^{\varepsilon},\dots,\mathbf{p}_{n}^{\varepsilon}\right) & := & -\sum_{i=1}^{n}\sum_{h=1}^{H}p_{i,h}^{\varepsilon}\cdot\ln p_{i,h}^{\varepsilon}.
\end{eqnarray*}
It is worth stressing that the errors $\varepsilon_{i}$ make the
problem (\ref{eq: consistency constraints GME}) under-conditioned
for a generic choice of $J$ and $n$, as the number of unknowns in
the distribution is larger than the number of constraints. For this
reason, a criterion (such as the entropy maximization) is needed in
order to select a single probability distribution among several ones
that satisfy both consistency and normalization constraints.

In order to compare different models represented by two distributions
$\mathbf{q}$ and $\mathbf{p}$ on a set $\Omega$, a generalization
of Shannon's entropy can be worthwhile. This can be achieved by introducing
the \emph{cross entropy} (or Kullback-Leibler divergence) from $\mathbf{q}$
to a dominated distribution $\mathbf{p}$, which is defined by 
\begin{equation}
\mathrm{D_{KL}}(\mathbf{p}||\mathbf{q}):=\sum_{\alpha\in\Omega}p_{\alpha}\cdot\log\left(\frac{p_{\alpha}}{q_{\alpha}}\right).\label{eq: KL divergence}
\end{equation}
Here, the symbol ``$:=$'' denotes the introduction of a mathematical
entity (in this case, the cross entropy function) through a definition.
We also stress out that (\ref{eq: KL divergence}) is defined even
when some of the components of $\mathbf{p}$ vanish; in other words,
it is defined in the whole ($|\Omega|-1$)-dimensional probability
symplex. In this regard, since events with zero probability do not
contribute to the cross entropy, we assume the identity $0\cdot\log0=0$.
A similar argument holds for (\ref{eq: GEM entropy}).

The cross entropy formalizes a notion of ``distance'' between $\mathbf{p}$
and an initial distribution $\mathbf{q}$. Whenever $\mathbf{q}$
is the uniform distribution $\mathbf{u}$ over a fixed finite set
$\Omega=\{\alpha_{1},\dots,\alpha_{n}\}$, one finds that $\mathrm{D_{KL}}(\mathbf{p}||\mathbf{u})=n-H(\mathbf{p})$,
where the entropy $H(\mathbf{p})$ associated with $\mathbf{p}$ is
$-\sum_{\alpha\in\Omega}p_{\alpha}\cdot\log p_{\alpha}$. Therefore,
the minimization of $\mathrm{D_{KL}}$ reduces to the maximum entropy
approach. For our purposes, the distribution $\mathbf{q}$ can be
regarded as a way to encode the prior information, and the cross entropy
provides us with an adequate tool to track the dynamics of the model.

Using the cross entropy as objective function to be minimized, one
can extend the GME to the \emph{Generalized Cross Entropy} (GCE) approach.
Introducing the notation $\mathbb{I}:=[0;1]$, this translates into
the minimization of the objective function 
\begin{eqnarray}
 &  & \underset{{\scriptstyle \mathbb{I}^{JK}\times\mathbb{I}^{nH}}}{\mathrm{argmin}}\mathrm{D_{KL}}\left(\left(\mathbf{p}^{\beta_{1}},\dots,\mathbf{p}^{\beta_{J}},\mathbf{p}_{1}^{\varepsilon},\dots,\mathbf{p}_{n}^{\varepsilon}\right)||\left(\mathbf{q}^{\beta_{1}},\dots,\mathbf{q}^{\beta_{J}},\mathbf{q}_{1}^{\varepsilon},\dots,\mathbf{q}_{n}^{\varepsilon}\right)\right)\nonumber \\
 & = & \underset{{\scriptstyle \mathbb{I}^{JK}\times\mathbb{I}^{nH}}}{\mathrm{argmin}}\left(\sum_{j=1}^{J}\sum_{k=1}^{K}p_{j,k}^{\beta}\ln\left(\frac{p_{j,k}^{\beta}}{q_{j,k}^{\beta}}\right)+\sum_{i=1}^{n}\sum_{h=1}^{H}p_{i,h}^{\varepsilon}\ln\left(\frac{p_{i,h}^{\varepsilon}}{q_{i,h}^{\varepsilon}}\right)\right)\label{eq: GCE objective function}
\end{eqnarray}
where the optimization is conducted with respect to the distribution
$(\mathbf{p}^{\beta_{1}},\dots,$ $\mathbf{p}^{\beta_{J}},\mathbf{p}_{1}^{\varepsilon},\dots,\mathbf{p}_{n}^{\varepsilon})$
under the same constraints (\ref{eq: consistency constraints GME})
and (\ref{eq: normalizations}).

Concrete applications of GME and GCE methods could be hard to implement:
a computationally expensive optimization procedure, along with a slow
processing of real-time data and the general shortcomings of ``global''
techniques, make their implementation often unfeasible. Furthermore,
on a conceptual ground, time effects get lost in these approaches.

\section{Characteristics and range of application of Stre-GCE}

\label{sec: Characteristics and range of application}

Before moving to the algorithm definition and the analysis of its
performance via simulations, we discuss the characteristics of the
Stre-GCE method in light of the issues mentioned in previous sections.

First, the consistency constraints and the temporal dimension can
be both preserved through the \emph{relaxation} of the GME method.
Specifically, the standard GCE with uniform prior involves the whole
set of constraints (\ref{eq: consistency constraints GME}), while
a time-sensitive acquisition of information follows from subsequent
upgrades of the distribution $(\mathbf{p}^{\beta_{1}},\dots,\mathbf{p}^{\beta_{J}})$
via a step-by-step introduction of constraints: only the $i^{\mathrm{th}}$
equation (\ref{eq: consistency constraints}) is strictly satisfied
at the $i^{\mathrm{th}}$ step. However, constraints at previous steps
affect the computation through the objective function: this leads
us to choose the cross entropy to move from a prior distribution to
a posterior one as new data show up.

It is worth to note that, in contrast with the standard GCE method,
our definition of the prior does not rely on information which is
external to the dataset: 
\begin{itemize}
\item we start from a \emph{uniform} distribution on the supports of parameters
and errors, and provide a first prior evaluation based on a batch
of initial data; 
\item the model is updated by incorporating new information to the prior.
This adaptive step involves the updating of both the estimates and
the prior. 
\end{itemize}
Therefore, one can see that the prior does not encode knowledge external
to the process, but it allows us to identify the temporal aspects
of data acquisition. To this extent, Stre-GCE is sensitive to the
time flow, hence it can be used when irreversibility plays a relevant
role.

Such a relaxation also leads to \emph{computational advantages}, as
it allows to move from a constrained problem with $n$ consistency
conditions to $n$ problems, each made up of a single condition. This
way, a solution to new problems can be found even when the original
problem is computationally unmanageable with the available softwares.
The splitting strategy can be modified in order to get intermediate
cases between Stre-GCE and standard GME/GCE: instead of individual
observations, we can consider the data received within a given time
frame. Processing partitions of the set of units $\{1,\dots,n\}$
in pairwise disjoint subsets $\mathcal{N}_{1},\dots,\mathcal{N}_{g}$,
e.g. intervals, such that $\bigcup_{\ell=1}^{g}\mathcal{N}_{\ell}=\{1,\dots,n\}$,
allows us to infer parameters and errors through consecutive applications
of GCE method within each individual time window. For $g=1$, one
has the original GCE problem with uniform prior, while at $g=n$ one
recovers the Stre-GCE. The steps in between will be referred to as
``\emph{block Stre-GCE}''.

Looking at the dual (or compressed) problem, the difference between
the two extreme solutions (GCE with uniform prior and Stre-GCE) can
be identified through the Lagrange multipliers that optimise the respective
objective functions. These multipliers allow the natural weights involved
in empirical likelihood techniques (see e.g. \cite{Golan1998}) to
change as the sequential inclusion of constraints progresses. On the
computational side, there are $J$ equations in the dual problem,
each of which is associated with a parameter $\beta_{j}$, $j\in\{1,\dots,J\}$.
In several applications, $J$ is smaller than the number of units
$n$ and, in any case, it is fixed for a given model. However, in
order to recover a solution from the minimization of the dual function
(that is the free energy or Legendre transform of the entropy), one
has to evaluate all the Lagrange multipliers associated with the constraints
(\ref{eq: consistency constraints GME}). Depending on the available
methods for the evaluation and the number of parameters and observations,
this task can still be difficult to accomplish. Moreover, such a formulation
cannot be applied exactly when one considers a more general class
of nonlinear (and, in particular, non-convex) optimization problems
when the equivalence of the two formulations (strong duality) does
not hold. Indeed, alternative methods and approximations are studied
to overcome this problem and get (pseudo-)optimal solutions \cite{Bertsekas1975}.

On the other hand, the Stre-GCE relaxation through subsequent updates
of estimations is manageable from both the primal and the dual perspectives.
Similarly to Recursive Least Squares (RLS) algorithms and the Kalman
Filter \cite{Simon2006}, these methods result into faster computations
as only partial information has to be managed at intermediate times.

Finally, we remark a conceptual shift from GCE with uniform prior
to Stre-GCE: we get an \emph{intrinsic} method to assign different
roles to signal and error. Other approaches usually claim for external
parameters to be added, whose evaluation is based on given optimization
criteria \cite{Wu2009}. On the contrary, Stre-GCE estimates the
different entropic contributions for both the signal and error as
they evolve in time, without the need of any additional variable:
in Stre-GCE, the prior for signal is sensitive to time evolution,
while the prior for error is identically uniform. Consequently, it
is possible to look at how the information evolves at intermediate
times in terms of the entropy production, which is a fundamental concept
in nonequilibrium processes \cite{Jarzynski1997,Crooks1999}. We
will focus on this issue in a separate work.

\section{Stre-GCE algorithm}

\label{sec: Stre-GCE algorithm}

The previous discussion can be formalized in the scheme pictured in
Figure \ref{fig: Stre-GCE scheme}. This is the basis for the Stre-GCE
algorithm, formally implemented in this section. 
\begin{center}
\begin{figure}[!ht]
\includegraphics[width=0.8\textwidth]{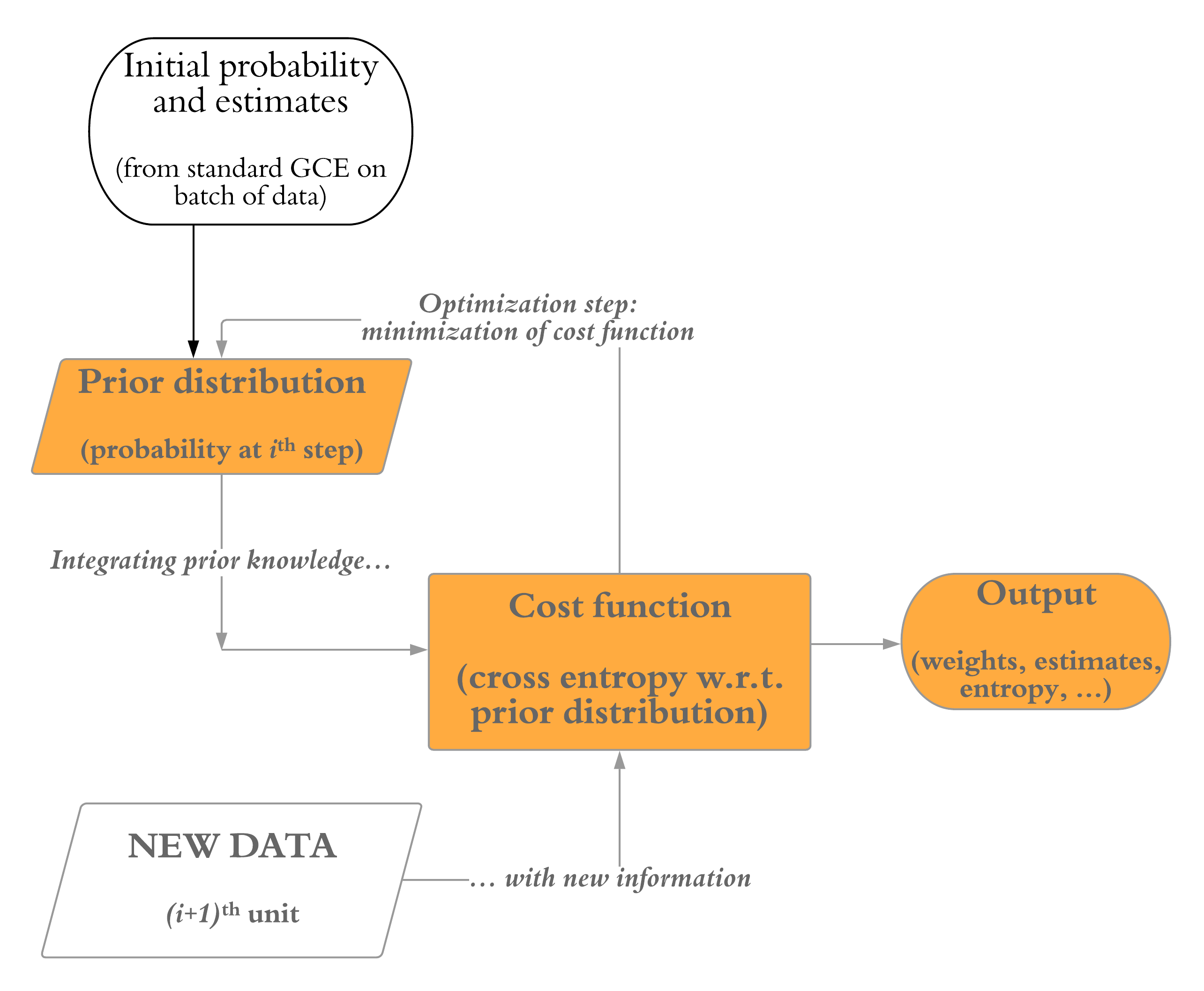} \caption{Scheme of the Stre-GCE algorithm.}
\label{fig: Stre-GCE scheme} 
\end{figure}
\par\end{center}

\subsubsection*{Initial step}

We start from an initial situation where a number $m$ of consistency
constraints, with $m<n$, have been collected. This provides us with
a \textit{batch} from which the initial estimates can be obtained.
Specifically, we can get 
\begin{eqnarray}
\hat{\beta}_{j}^{(m)} & := & \sum_{k=1}^{K}\hat{p}_{j,k}^{\beta(m)}\cdot z_{j,k}^{\beta},\quad j\in\{1,\dots,J\},\nonumber \\
\hat{\varepsilon}_{i} & := & \sum_{h=1}^{H}\hat{p}_{i,h}^{\varepsilon}\cdot z_{i,h}^{\varepsilon},\quad i\in\{1,\dots,m\}\label{eq: base step first measurement}
\end{eqnarray}
through the GCE method with uniform prior. As already mentioned, this
choice of the prior leads to the equivalence with the standard GME
from the optimization view, so this approach selects the probability
distribution 
\begin{eqnarray}
(\mathbf{\hat{P}}^{\beta(m)},\mathbf{\hat{p}}_{1}^{\varepsilon},\dots,\mathbf{\hat{p}}_{m}^{\varepsilon}) & := & \underset{{\scriptstyle \mathbb{I}^{JK}\times\mathbb{I}^{mH}}}{\mathrm{argmin}}\mathrm{D_{KL}}\left((\mathbf{P}^{\beta},\mathbf{p}_{1}^{\varepsilon},\dots,\mathbf{p}_{m}^{\varepsilon})||(\mathbf{U}^{\beta},\mathbf{u}^{\varepsilon},\dots,\mathbf{u}^{\varepsilon})\right)\nonumber \\
 & = & \underset{{\scriptstyle \mathbb{I}^{JK}\times\mathbb{I}^{mH}}}{\mathrm{argmax}}\left(H(\mathbf{P}^{\beta})+\sum_{i=1}^{m}H(\mathbf{p}_{i}^{\varepsilon})\right)\nonumber \\
 & = & \underset{{\scriptstyle \mathbb{I}^{JK}\times\mathbb{I}^{mH}}}{\mathrm{argmax}}\left(-\sum_{j=1}^{J}\sum_{k=1}^{K}p_{j,k}^{\beta}\cdot\ln p_{j,k}^{\beta}-\sum_{h=1}^{H}\sum_{i=1}^{m}p_{i,h}^{\varepsilon}\cdot\ln p_{i,h}^{\varepsilon}\right)\nonumber \\
\label{eq: base step, estimate}
\end{eqnarray}
where the choice set is the set of all the distributions $(\mathbf{P}^{\beta},\mathbf{p}_{1}^{\varepsilon},\dots,\mathbf{p}_{m}^{\varepsilon})$
satisfying the constraints 
\begin{equation}
\left\{ \begin{array}{c}
{\displaystyle y_{i}=\sum_{j=1}^{J}\sum_{k=1}^{K}p_{j,k}^{\beta}\cdot z_{j,k}^{\beta}\cdot x_{i,j}+\sum_{h=1}^{H}p_{i,h}^{\varepsilon}z_{i,h}^{\varepsilon}},\quad i\in\{1,\dots,m\}\\
{\displaystyle \sum_{k=1}^{K}p_{j,k}^{\beta}=\sum_{h=1}^{H}p_{i,h}^{\varepsilon}=1,\quad j\in\{1,\dots,J\},\,i\in\{1,\dots,m\}}
\end{array}\right..\label{eq: base step, constraints}
\end{equation}
From the method of Lagrange multipliers associated with these linear
constraints, one gets the weights of Gibbs' type (see e.g. \cite{LL1980,Golan2018}
for more details) 
\begin{eqnarray}
\hat{p}_{j,k}^{\beta(m)} & = & \frac{1}{Z_{j}^{\beta(m)}}\cdot\exp\left(-z_{j,k}^{\beta}\cdot\sum_{i=1}^{m}x_{i,j}\hat{\lambda}_{i}\right),\quad j\in\{1,\dots,J\},\,k\in\{1,\dots,K\},\nonumber \\
\hat{p}_{i,h}^{\varepsilon} & = & \frac{1}{Z_{i}^{\varepsilon}}\cdot\exp\left(-z_{i,h}^{\varepsilon}\hat{\lambda}_{i}\right),\quad i\in\{1,\dots,m\},\,h\in\{1,\dots,H\}.\label{eq: Gibbs' weights}
\end{eqnarray}
where the multipliers $\hat{\lambda}_{i}$ are chosen in order to
get (\ref{eq: consistency constraints}) and the normalizing factors
$Z_{j}^{\beta(m)}$, $Z_{i}^{\varepsilon}$ ensure that (\ref{eq: normalizations})
holds.

\subsubsection*{Update step: $i\rightarrow i+1$}

In the update process, the actual estimates for signal and error are combined with the new observation in order to produce a new estimate. This purpose
is attained through the minimization of the cross entropy (or Kullback-Leibler
divergence) $\mathrm{D_{KL}}((\mathbf{P}_{i+1}^{\beta},\mathbf{p}_{i+1}^{\varepsilon})||(\mathbf{\hat{P}}_{i}^{\beta},\mathbf{u}_{i}^{\varepsilon}))$.
The asymmetry between signal and error is naturally expressed by different
priors for $\mathbf{P}^{\beta}$ and $\mathbf{p}^{\varepsilon}$:
in fact, the evolution of $\mathbf{P}^{\beta}$ from the $i^{\mathrm{th}}$
to the $(i+1)^{\mathrm{th}}$ step depends on the prior $\mathbf{\hat{P}}_{i}^{\beta}$
at time $i$, while $\mathbf{p}^{\varepsilon}$ does not \textit{directly}
depend on previous steps. This assumption is expressed choosing the
uniform distribution on the support $\left\{ z_{i+1,1}^{\varepsilon},\dots,z_{i+1,H}^{\varepsilon}\right\} $
for the prior $\mathbf{u}_{i}^{\varepsilon}$: this choice for the
update rule allows to distinguish between signal and errors in terms
of effects of the prior knowledge about them on the knowledge at subsequent
times. The asymmetry is further stressed by the estimates obtained
from the adaptive process by $\mathbf{\hat{P}}$, in contrast to the
estimates that are not sensitive to time effects, such as the uniform
distributions $\mathbf{u}_{i}^{\varepsilon}$.

Apart from a known constant term $\ln H$, which does not affect the
optimization step, the objective function can be expressed as 
\begin{equation}
\mathrm{D_{KL}}(\mathbf{P}^{\beta}||\mathbf{\hat{P}}_{i}^{\beta})-H(\mathbf{p}^{\varepsilon})\label{eq: KL objective function}
\end{equation}
in order to get 
\begin{eqnarray}
 &  & (\mathbf{\hat{P}}_{i+1}^{\beta},\mathbf{\hat{p}}_{i+1}^{\varepsilon})\nonumber \\
 & := & \underset{{\scriptstyle \mathbb{I}^{JK}\times\mathbb{I}^{H}}}{\mathrm{argmin}}\mathrm{D_{KL}}\left((\mathbf{P}^{\beta},\mathbf{p}_{i+1}^{\varepsilon})||(\mathbf{\hat{P}}_{i}^{\beta},\mathbf{u}_{i}^{\varepsilon})\right)\nonumber \\
 & = & \underset{{\scriptstyle \mathbb{I}^{JK}\times\mathbb{I}^{H}}}{\mathrm{argmin}}\left(\sum_{j=1}^{J}\sum_{k=1}^{K}p_{j,k}^{\beta}\ln p_{j,k}^{\beta}-\sum_{j=1}^{J}\sum_{k=1}^{K}p_{j,k}^{\beta}\ln\hat{p}_{j,k}^{\beta(i)}+\sum_{h=1}^{H}p_{i+1,h}^{\varepsilon}\ln p_{i+1,h}^{\varepsilon}\right)\nonumber \\
\label{eq: update step, estimates}
\end{eqnarray}
under the requirements 
\begin{equation}
\left\{ \begin{array}{c}
y_{i+1}={\displaystyle \sum_{j=1}^{J}\sum_{k=1}^{K}p_{j,k}^{\beta}\cdot z_{j,k}^{\beta}\cdot x_{j,i+1}+\sum_{h=1}^{H}p_{i+1,h}^{\varepsilon}z_{i+1,h}^{\varepsilon}}\\
{\displaystyle \sum_{k=1}^{K}p_{j,k}^{\beta}=\sum_{h=1}^{H}p_{i+1,h}^{\varepsilon}=1,\quad j\in\{1,\dots,J\}}
\end{array}\right.\label{eq: update step, constraints}
\end{equation}
for the candidate solutions $(\mathbf{P}_{i+1}^{\beta},\mathbf{p}_{i+1}^{\varepsilon})$.

The relaxation mentioned in Section \ref{sec: Characteristics and range of application}
is manifest as it can be seen by comparing (\ref{eq: base step, constraints})
and (\ref{eq: update step, constraints}). With regard to the dual
problem, the analytic expression of the solutions of (\ref{eq: update step, estimates})
is still in Gibbs' form, since at each step the prior $\mathbf{\hat{p}}_{i}^{\beta}$
are in Gibbs' form and the constraints are linear.

It is worth remarking that the temporal effects are represented by
the cross term $-\sum_{j=1}^{J}\sum_{k=1}^{K}p_{j,k}^{\beta}\ln\hat{p}_{j,k}^{\beta(i)}$
in the objective function. Even if (\ref{eq: KL objective function})
does not include a dependence on $\mathbf{\hat{p}}_{t}^{\varepsilon}$
for $t\leq i$, these distributions are related to the prior $\mathbf{\hat{P}}_{i}^{\beta}$
through the consistency constraints. Hence, they indirectly appear
in the optimization process, even if on a different footing with respect
to $\mathbf{P}^{\beta}$.

\section{Simulation study}

\label{sec: Simulations}

In order to test the efficiency of the method, we performed several
simulations in GAMS. We defined a set of simulations to compare the
GCE method with uniform prior on the whole dataset, the GCE only on
the batch, and the Stre-GCE. Their performance has been evaluated
through Root-Mean-Square Error (RMSE) 
\begin{equation}
\mathrm{RMSE}:=\sqrt{\frac{1}{n}\cdot\sum_{i=1}^{n}(y_{i}-\boldsymbol{\hat{\beta}}\cdot\mathbf{x}_{i})^{2}}.\label{eq: rmse}
\end{equation}
to quantify how large is the ``distance'' between the solutions
for the standard GCE with uniform prior and the Stre-GCE. 
RMSEs resulting from the simulations have been used for evaluation of the different methods, as they allow to compare the corresponding estimates based on the same sample. In particular, for
each dataset $\mathcal{D}$ under consideration, we evaluated (\ref{eq: rmse})
using the estimates $\boldsymbol{\hat{\beta}}$ provided by standard
GCE and Stre-GCE, respectively denoted as $\mathrm{RMSE}_{\text{GCE}}(\mathcal{D})$
and $\mathrm{RMSE}_{\text{Stre-GCE}}(\mathcal{D})$.
Then, we focused on the ratio 
\begin{equation}
\varrho:=\frac{\mathrm{RMSE}_{\text{Stre-GCE}}(\mathcal{D})}{\mathrm{RMSE}_{\text{GCE}}(\mathcal{D})}.\label{eq: efficiency}
\end{equation}
This ratio, or equivalently the value $\varrho-1$, describes the
efficiency of the streaming approach compared to the standard one. 

To further explore the dependence of Stre-GCE on the dataset, we also
checked the algorithm when variables are standardized and when multicollinearity
effects occur: we considered the performance on both the sampled independent
variables, which give $\mathbf{y}$ via (\ref{eq: matrix consistency constraint}),
and on their standardized version. The degree of multicollinearity
is simulated considering the following combination
of standardized variables 
\begin{equation}
x_{j}\mapsto\eta\cdot c+\sqrt{1-\eta^{2}}\cdot x_{j},\label{eq: degree multicollinearity}
\end{equation}
so one gets perfect multicollinearity at $\eta=1$. The
efficiency (\ref{eq: efficiency}) in case of standardized variables
and multicollinearity has been denoted by $\varrho_{\text{STD}}$
and $\varrho_{\eta}$ respectively. 

We chose $J=3$ and fixed the value of the intercept at $\alpha=1$,
so $\alpha$ will not impact on performance evaluation of the method. Several tests were run at various sample sizes and batch ratios 
\begin{equation}
r:=\frac{\text{batch size}}{\text{full sample size}}.
\label{eq: batch ratio}
\end{equation}
We considered samples of sizes $n\in\left\{ 60,120,240,480,960,1920,3840\right\} $,
while we mainly focused on batches of dimensions $\frac{1}{4}n$,
$\frac{1}{2}n$ and $\frac{3}{4}n$ to get the initial estimates for
$\mathbf{p}^{\beta}$. 

The variables $x_{i,j}$, as long as $c$ in (\ref{eq: degree multicollinearity}),
were sampled from a uniform distribution over in $[0,20]$, while
errors $\varepsilon_{i}$ came from the normal distribution $\mathcal{N}(0,1)$.
The supports for the estimates of the signal and the error are 
\begin{equation}
\left\{ z_{j,k}^{\beta}:\,k\in\{1,\dots,K\}\right\} =\left\{ -100,-50,0,50,100\right\} \quad j\in\{1,\dots,J\}\label{eq: support signal}
\end{equation}
and 
\begin{equation}
\left\{ z_{i,h}^{\varepsilon}:\,h\in\{1,\dots,H\}\right\} =\left\{ -3\cdot s_{y},0,3\cdot s_{y}\right\} ,\quad i\in\{1,\dots,n\}\label{eq: support errors}
\end{equation}
respectively. The sample standard deviation $s_{y}=\sqrt{\frac{1}{n-1}\sum_{i=1}^{n}(y_{i}-\overline{y})^{2}}$
was used in (\ref{eq: support errors}) in accordance to the $3$-sigma
rule \cite{Pukelsheim1994,Golan1996}, in order to make the scale
of support compatible with $s_{y}$.

Finally, based on the observations in Section \ref{sec: Characteristics and range of application},
we extended our algorithm in order to include the dimension $g$ of
blocks, specifically we look at $g\in\{1,10,20,40\}$.

\subsubsection*{First set of simulations: different sizes and batches}

The focus of this set of simulations is the comparison between GCE
on the whole sample, GCE on the batch, and Stre-GCE starting from
the estimate of the batch. The efficiencies $\varrho$
and $\varrho_{\mathrm{STD}}$ are shown in Table \ref{tab: rmse at sample sizes 60, 120, 240, 480, 960, 1920, 3840}: starting from the $\mathrm{RMSE}_{\text{GCE}}$ associated with each dataset, it is possible to recover the original $\mathrm{RMSE}_{\text{Stre-GCE}}$ (\ref{eq: efficiency}). For the sake of completeness, we also include the $\mathrm{RMSE}_{\text{GCE}}$ applied only to the batch. 

\begin{table}
\begin{centering}
\begin{tabular}{|c|c|c|c|c|}
\hline
Sample size $n$                                       & Batch ratio $r$            & $\mathrm{RMSE}_{\text{GCE on batch}}$ & $\varrho$ & $\varrho_{\mathrm{STD}}$ \\ \hline
$n=60$                                        & $\nicefrac{1}{4}$ & 0,7957                                & 1,0594    & 1,2079                   \\ \cline{2-5} 
$\mathrm{RMSE}_{\text{GCE}}$=1,0702       & $\nicefrac{2}{4}$ & 0,8946                                & 1,1166    & 1,2077                   \\ \cline{2-5} 
$\mathrm{RMSE}_{\text{GCE (STD)}}$=1,0745 & $\nicefrac{3}{4}$ & 0,9908                                & 1,0490    & 1,1932                   \\ \hline
$n=120$                                       & $\nicefrac{1}{4}$ & 1,1064                                & 1,1329    & 1,0557                   \\ \cline{2-5} 
$\mathrm{RMSE}_{\text{GCE}}$=1,0494       & $\nicefrac{2}{4}$ & 1,0345                                & 1,1317    & 1,0557                   \\ \cline{2-5} 
$\mathrm{RMSE}_{\text{GCE (STD)}}$=1,0504 & $\nicefrac{3}{4}$ & 1,0448                                & 1,0601    & 1,0558                   \\ \hline
$n=240$                                       & $\nicefrac{1}{4}$ & 1,0223                                & 1,0712    & 1,0240                   \\ \cline{2-5} 
$\mathrm{RMSE}_{\text{GCE}}$=1,0062       & $\nicefrac{2}{4}$ & 1,0512                                & 1,0732    & 1,0240                   \\ \cline{2-5} 
$\mathrm{RMSE}_{\text{gCE (STD)}}$=1,0065 & $\nicefrac{3}{4}$ & 1,0355                                & 1,0733    & 1,0240                   \\ \hline
$n=480$                                       & $\nicefrac{1}{4}$ & 0,8614                                & 1,1497    & 1,3190                   \\ \cline{2-5} 
$\mathrm{RMSE}_{\text{GCE}}$=0,9133       & $\nicefrac{2}{4}$ & 0,9192                                & 1,1360    & 1,3190                   \\ \cline{2-5} 
$\mathrm{RMSE}_{\text{gCE (STD)}}$=0,9134 & $\nicefrac{3}{4}$ & 0,9268                                & 1,1337    & 1,3190                   \\ \hline
$n=960$                                       & $\nicefrac{1}{4}$ & 0,9518                                & 1,0721    & 1,1049                   \\ \cline{2-5} 
$\mathrm{RMSE}_{\text{GCE}}$=0,9976       & $\nicefrac{2}{4}$ & 1,0048                                & 1,0739    & 1,1049                   \\ \cline{2-5} 
$\mathrm{RMSE}_{\text{gCE (STD)}}$=0,9976 & $\nicefrac{3}{4}$ & 0,9808                                & 1,0751    & 1,1049                   \\ \hline
$n=1920$                                      & $\nicefrac{1}{4}$ & 0,9649                                & 2,8940    & 1,6075                   \\ \cline{2-5} 
$\mathrm{RMSE}_{\text{GCE}}$=1,0072       & $\nicefrac{2}{4}$ & 0,9980                                & 2,8872    & 1,6075                   \\ \cline{2-5} 
$\mathrm{RMSE}_{\text{gCE (STD)}}$=1,0101 & $\nicefrac{3}{4}$ & 1,0129                                & 2,9034    & 1,6075                   \\ \hline
$n=3840$                                      & $\nicefrac{1}{4}$ & 0,9995                                & 1,6790    & 1,0448                   \\ \cline{2-5} 
$\mathrm{RMSE}_{\text{GCE}}$=1,0066       & $\nicefrac{2}{4}$ & 1,0139                                & 1,6767    & 1,0448                   \\ \cline{2-5} 
$\mathrm{RMSE}_{\text{gCE (STD)}}$=0,9979 & $\nicefrac{3}{4}$ & 1,0112                                & 1,6730    & 1,0448                   \\ \hline
\end{tabular}
\par\end{centering}
\caption{Efficiencies $\varrho$ and $\varrho_{\text{STD}}$ for Stre-GCE at different values of sample size $n$ and batch ratio $r$. The RMSE associated with the GCE method applied on the batch has been reported too.}
\label{tab: rmse at sample sizes 60, 120, 240, 480, 960, 1920, 3840} 
\end{table}

We note that the behavior of the performance of Stre-GCE is not monotonic
with respect to the size of the dataset: this also holds for standard
GCE, as it can be seen moving from $n=480$ to $n=960$ 
or $n=1920$. In Stre-GCE, the dependence on the specific dataset is amplified by the time ordering of the data during the stream: units at different time steps have different roles, so they may give rise to fluctuations in the performance, which may be suppressed by new data over time.

\subsubsection*{Second set of simulations: stream of blocks}

The \emph{streaming blocks updates} can be considered to soften these
oscillations.

We chose the block dimensions $g\in\{1,10,20,40\}$ and considered a different
seed for simulations in order to extend the exploration of the performance
of Stre-GCE beyond the datasets considered in the previous paragraph. The block dimension $g=1$ corresponds to the standard Stre-GCE method described above. The results are set out in
Table \ref{tab: rmse for streaming blocks at sample sizes 480, 960}, where the ratios 
\begin{equation}
\varrho_{\text{block},g}:=\frac{\mathrm{RMSE}_{\text{Streaming Block GCE}}(\mathcal{D})}{\mathrm{RMSE}_{\text{GCE}}(\mathcal{D})}\label{eq: efficiency blocks}
\end{equation}
are shown for sample sizes $n\in\{120,240,480,960,1920,3840\}$ and batch ratios $r\in\{\nicefrac{1}{4},\nicefrac{2}{4},\nicefrac{3}{4}\}$.

\begin{table}
\begin{centering}
\begin{tabular}{|c|c|c|c|c|c|}
\hline
Sample size $n$                     & Batch ratio $r$   & $\varrho_{\text{block},1}$ & $\varrho_{\text{block},10}$ & $\varrho_{\text{block},20}$ & $\varrho_{\text{block},40}$ \\ \hline
$n=120$                             & $\nicefrac{1}{4}$ & 1,0423                     & 1,0284                      & 1,0247                      & 1,0329                      \\ \cline{2-6} 
$\mathrm{RMSE}_{\text{GCE}}$=0,9146 & $\nicefrac{2}{4}$ & 1,0039                     & 1,0281                      & 1,0163                      & 1,0145                      \\ \cline{2-6} 
                                    & $\nicefrac{3}{4}$ & 1,0517                     & 1,0233                      & 1,0233                      & 1,0366                      \\ \hline
$n=240$                             & $\nicefrac{1}{4}$ & 1,4349                     & 1,0257                      & 1,0623                      & 1,0626                      \\ \cline{2-6} 
$\mathrm{RMSE}_{\text{GCE}}$=0,9179 & $\nicefrac{2}{4}$ & 1,4296                     & 1,0257                      & 1,0623                      & 1,0343                      \\ \cline{2-6} 
                                    & $\nicefrac{3}{4}$ & 1,4256                     & 1,0257                      & 1,0623                      & 1,0627                      \\ \hline
$n=480$                             & $\nicefrac{1}{4}$ & 1,0306                     & 1,0399                      & 1,0210                      & 1,0147                      \\ \cline{2-6} 
$\mathrm{RMSE}_{\text{GCE}}$=1,0097 & $\nicefrac{2}{4}$ & 1,0786                     & 1,0399                      & 1,0210                      & 1,0147                      \\ \cline{2-6} 
                                    & $\nicefrac{3}{4}$ & 1,0812                     & 1,0398                      & 1,0209                      & 1,0145                      \\ \hline
$n=960$                             & $\nicefrac{1}{4}$ & 1,2441                     & 1,0544                      & 1,0589                      & 1,0920                      \\ \cline{2-6} 
$\mathrm{RMSE}_{\text{GCE}}$=1,0268 & $\nicefrac{2}{4}$ & 1,2526                     & 1,0544                      & 1,0589                      & 1,0920                      \\ \cline{2-6} 
                                    & $\nicefrac{3}{4}$ & 1,2553                     & 1,0544                      & 1,0589                      & 1,0920                      \\ \hline
$n=1920$                            & $\nicefrac{1}{4}$ & 1,5049                     & 1,0777                      & 1,0924                      & 1,0301                      \\ \cline{2-6} 
$\mathrm{RMSE}_{\text{GCE}}$=0,9986 & $\nicefrac{2}{4}$ & 1,5474                     & 1,0777                      & 1,0924                      & 1,0301                      \\ \cline{2-6} 
                                    & $\nicefrac{3}{4}$ & 1,5665                     & 1,0777                      & 1,0924                      & 1,0301                      \\ \hline
$n=3840$                            & $\nicefrac{1}{4}$ & 1,0568                     & 1,1128                      & 1,0169                      & 1,0060                      \\ \cline{2-6} 
$\mathrm{RMSE}_{\text{GCE}}$=0,9738 & $\nicefrac{2}{4}$ & 1,0568                     & 1,1128                      & 1,0169                      & 1,0060                      \\ \cline{2-6} 
                                    & $\nicefrac{3}{4}$ & 1,0561                     & 1,1128                      & 1,0169                      & 1,0060                      \\ \hline
\end{tabular}
\par\end{centering}
\caption{Values of $\varrho_{\text{block},g}$ for block Stre-GCE at block dimension
$g\in\{1,10,20,40\}$, datasets with sizes $n\in\{120,240,480,960,1920,3840\}$ and batch ratios $r\in\{\nicefrac{1}{4},\nicefrac{2}{4},\nicefrac{3}{4}\}$.}
\label{tab: rmse for streaming blocks at sample sizes 480, 960} 
\end{table}

\subsubsection*{Third set of simulations: multicollinearity}

Different degrees of multicollinearity were simulated: we chose $\eta\in\{0.2,0.4,$
$0.6,0.8,1\}$ so that we range from multicollinearity absence ($\eta=0$)
to perfect multicollinearity ($\eta=1$). Results are reported in
Table \ref{tab: rmse for multicollinearity at sample sizes 240, 480, 960}
for samples of sizes 120, 240, 480, 960 and 1920. Only results with batch ratio $r=\nicefrac{1}{4}$ are reported. In fact, except for $\varrho_{\eta}=1,0774
	$ at $(n;\eta;r)=(120;0,6;\nicefrac{3}{4})$, $\varrho_{\eta}=1,0477$
	at $(n;\eta;r)=(120;0,8;\nicefrac{3}{4})$, and  $\varrho_{\eta}=1,0071$
	at $(n;\eta;r)=(1920;1;\nicefrac{2}{4})$, no major difference (i.e. less than $10^{-3}$) was found for the efficiency $\varrho_{\eta}$ at fixed dataset, fixed $\eta$ and varying the batch ratio in $\{\nicefrac{1}{4},\nicefrac{2}{4},\nicefrac{3}{4}\}$.

\begin{table}
\begin{centering}
\begin{tabular}{|c|c|c|c|c|c|c|}
\hline
\multicolumn{2}{|c|}{Sample size $n$} & $\eta=0.2$ & $\eta=0.4$ & $\eta=0.6$ & $\eta=0.8$ & $\eta=1$ \\ \hline
$n=120$   & $\mathrm{RMSE}_{\text{GCE}}$  & 1,0505     & 1,0507     & 1,0510     & 1,0521     & 1,0600   \\ \cline{2-7}
      & $\varrho_{\eta}$              & 1,0463     & 1,0586     & 1,0819     & 1,0649     & 1,0271   \\ \hline
$n=240$   & $\mathrm{RMSE}_{\text{GCE}}$  & 1,0081     & 1,0101     & 1,0117     & 1,0128     & 1,0168   \\ \cline{2-7} 
      & $\varrho_{\eta}$              & 1,0549     & 1,1016     & 1,1753     & 1,2691     & 1,0175   \\ \hline
$n=480$   & $\mathrm{RMSE}_{\text{GCE}}$  & 0,9144     & 0,9152     & 0,9155     & 0,9154     & 0,9170   \\ \cline{2-7} 
      & $\varrho_{\eta}$              & 1,2686     & 1,2237     & 1,1771     & 1,1203     & 1,1968   \\ \hline
$n=960$   & $\mathrm{RMSE}_{\text{GCE}}$  & 0,9969     & 0,9959     & 0,9952     & 0,9948     & 0,9950   \\ \cline{2-7} 
      & $\varrho_{\eta}$              & 1,0929     & 1,2853     & 1,6471     & 1,8900     & 1,6000   \\ \hline
$n=1920$  & $\mathrm{RMSE}_{\text{GCE}}$  & 1,0063     & 1,0064     & 1,0064     & 1,0065     & 1,0071   \\ \cline{2-7} 
      & $\varrho_{\eta}$              & 1,3546     & 1,6324     & 1,5652     & 1,4270     & 1,3546   \\ \hline
\end{tabular}
\par\end{centering}
\caption{Efficiency $\varrho_{\eta}$ for Stre-GCE when multicollinearity (expressed by $\eta$) occurs,
at different values of the sample size and batch dimensions. Comparison
with GCE on the whole dataset is highlighted. }
\label{tab: rmse for multicollinearity at sample sizes 240, 480, 960} 
\end{table}

The previous simulations show an evident distortion between the usual
GCE method and the results of Stre-GCE. The extent of this distortion
statistically depends on the sample, but a structural contribution
comes from the consolidation of fluctuations and errors at each step
of the iteration, which results in a structural bias at the end of
the process. Indeed, the Stre-GCE approach described in Section \ref{sec: Stre-GCE algorithm}
is highly sensitive to the inclusion of new data: a single datum associated
with large error $\varepsilon_{i}$ may reduce the quality of the
estimate due to consistency constraint (\ref{eq: update step, constraints}).
This means that the quality of the final estimate may show considerable
fluctuations as the data dimension increases.

In the following section, we will discuss an extension of the Stre-GCE
algorithm aimed at including the role of ``consolidated'' knowledge.

\section{Balancing experience and new data: Weighted Stre-GCE }

\label{sec: Weighted Stre-GCE} In this section, we further tune the
balance between entropy contributions due to the signal and the error.
These two contributions appear on an equal footing in the Stre-GCE
algorithm. On the other hand, time may also affect the \emph{weights}
with which the two entropy contributions are combined. For instance,
assuming that an underlying model (e.g., the parameters $\beta$)
does not change during the data stream, the ``plausibility'' of
consolidated information increases if compared with new data.

This argument can be formalized in the generalization of the Stre-GCE
objective function (\ref{eq: KL objective function}) with the inclusion
of weights 
\begin{equation}
w_{\beta}^{(i)}\cdot\mathrm{D_{KL}}(\mathbf{P}^{\beta}||\mathbf{\hat{P}}_{i}^{\beta})-w_{\varepsilon}^{(i)}\cdot H(\mathbf{p}^{\varepsilon})\label{eq: weighted KL objective function}
\end{equation}
where $w_{\beta}^{(i)},w_{\varepsilon}^{(i)}\in\mathbb{I}$, $w_{\beta}^{(i)}\neq0$
and $w_{\beta}^{(i)}+w_{\varepsilon}^{(i)}=1$ for all $i$ indexing
the streaming data. Weights depend on time through the labelling $i$
and they enter the streaming updates through the solutions of the
optimization problems (\emph{Weighted Stre-GCE}) 
\begin{equation}
(\mathbf{\hat{P}}_{i+1}^{\beta},\mathbf{\hat{p}}_{i+1}^{\varepsilon}):=\underset{{\scriptstyle \mathbb{I}^{JK}\times\mathbb{I}^{H}}}{\mathrm{argmin}}\left(w_{\beta}^{(i)}\cdot\mathrm{D_{KL}}(\mathbf{P}^{\beta}||\mathbf{\hat{P}}_{i}^{\beta})-w_{\varepsilon}^{(i)}\cdot H(\mathbf{p}^{\varepsilon})\right)\label{eq: weighted update step, estimates}
\end{equation}
subject to the constraints (\ref{eq: update step, constraints}).
The same solution $(\mathbf{\hat{P}}_{i+1}^{\beta},\mathbf{\hat{p}}_{i+1}^{\varepsilon})$
is obtained when the objective function (\ref{eq: weighted KL objective function})
is multiplied by an arbitrary positive constant. In particular, we
will consider the scaled function 
\begin{equation}
\mathrm{D_{KL}}(\mathbf{P}^{\beta}||\mathbf{\hat{P}}_{i}^{\beta})-\frac{w_{\varepsilon}^{(i)}}{w_{\beta}^{(i)}}\cdot H(\mathbf{p}^{\varepsilon})\label{eq: weighted KL objective function, scaled}
\end{equation}
to assess the temporal dependence of the weights $w_{\beta}^{(i)},w_{\varepsilon}^{(i)}$.
In this work, we denote 
\begin{equation}
\omega^{(i)}:=\frac{w_{\varepsilon}^{(i)}}{w_{\beta}^{(i)}}\label{eq: weighting ratio}
\end{equation}
and choose the update rule 
\begin{equation}
\omega^{(i+1)}:=\frac{\omega^{(i)}}{\omega^{(i)}+1}.\label{eq: weighting ratio, update rule}
\end{equation}
The rationale behind this choice is easily described: let us denote
by $f_{\beta}^{(i)},f_{\varepsilon}^{(i)},f_{S}^{(i)}$ the ``\emph{dimensionalities}''
attributed to the signal, the error, and the whole sample, respectively.
When a new datum arrives, i.e. $i\mapsto i+1$, we ascribe its additional
contribution to the signal's dimensionality, leaving errors' dimensionality
unchanged. This assumption can be formalized in the following update
rule at $i\mapsto i+1$: 
\begin{align*}
f_{\beta}^{(i+1)}:= & f_{\beta}^{(i)}+1,\\
f_{\varepsilon}^{(i+1)}:= & f_{\varepsilon},\\
f_{S}^{(i+1)}:= & f_{S}^{(i)}+1.
\end{align*}
In particular, the dimensionality $f_{\varepsilon}$ is constant,
so we can omit the index $i$. We now express the weights $w_{\beta}^{(i)},w_{\varepsilon}^{(i)}$
in the Weighted Stre-GCE objective function (\ref{eq: weighted update step, estimates})
as frequencies related to the dimensionalities $f_{\beta}^{(i)},f_{\varepsilon}^{(i)},f_{S}^{(i)}$:
\begin{align*}
w_{\beta}^{(i+1)}:= & \frac{f_{\beta}^{(i+1)}}{f_{S}^{(i+1)}}=\frac{f_{\beta}^{(i)}+1}{f_{S}^{(i)}+1},\\
w_{\varepsilon}^{(i+1)}:= & \frac{f_{\varepsilon}^{(i+1)}}{f_{S}^{(i+1)}}=\frac{f_{\varepsilon}^{(i)}}{f_{S}^{(i)}+1}.
\end{align*}
Note that if the normalization condition is satisfied at the beginning
of the process, i.e. $w_{\beta}^{(0)}+w_{\varepsilon}^{(0)}=1$, then
it is preserved at subsequent times too. Then, the weighting ratio
(\ref{eq: weighting ratio}) changes as follows: 
\begin{align*}
\omega^{(i+1)}= & \frac{f_{\varepsilon}^{(i)}}{f_{\beta}^{(i)}+1},\\
= & \left(\frac{1}{\omega^{(i)}}+\frac{1}{f_{\varepsilon}}\right)^{-1}\\
= & \frac{\omega^{(i)}\cdot f_{\varepsilon}}{\omega^{(i)}+f_{\varepsilon}}.
\end{align*}
Choosing $f_{\varepsilon}:=1$, we recover (\ref{eq: weighting ratio, update rule}).
It is worth remarking that the recursive relation (\ref{eq: weighting ratio, update rule})
allows the update of weights without the explicit knowledge of the
dimension $n$ of the full sample, which is consistent with the streaming
approach under consideration.

\begin{figure}[ht]   
\begin{minipage}[b]{0.48\linewidth}
  	\centering
  	\includegraphics[width=\linewidth]{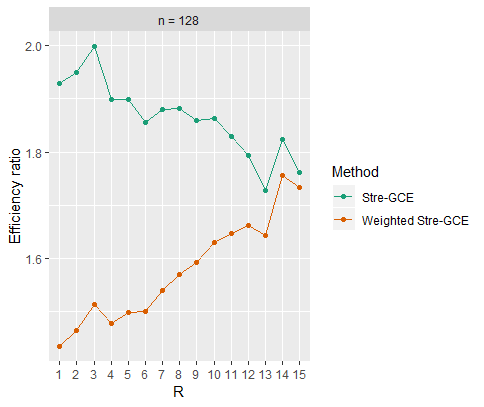} 
\end{minipage}
  \begin{minipage}[b]{0.48\linewidth}
    \centering
    \includegraphics[width=\linewidth]{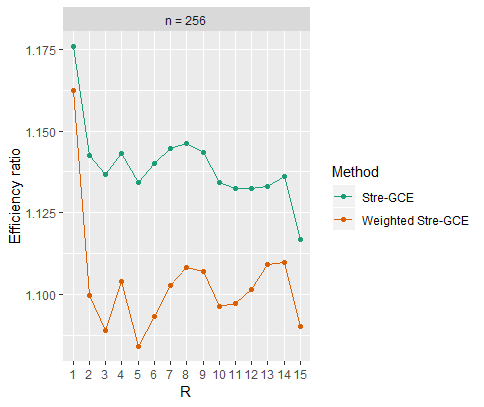} 
  \end{minipage} 
  \begin{minipage}[b]{0.48\linewidth}
    \centering
    \includegraphics[width=\linewidth]{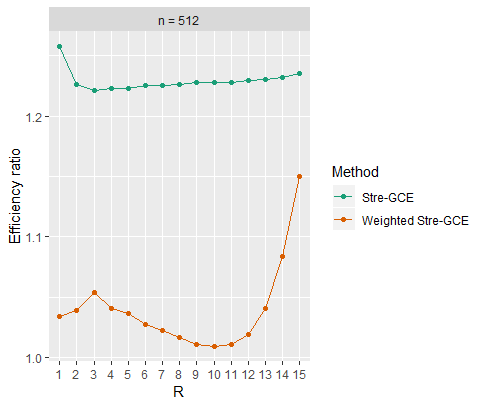} 
  \end{minipage} 
  \begin{minipage}[b]{0.48\linewidth}
    \centering
    \includegraphics[width=\linewidth]{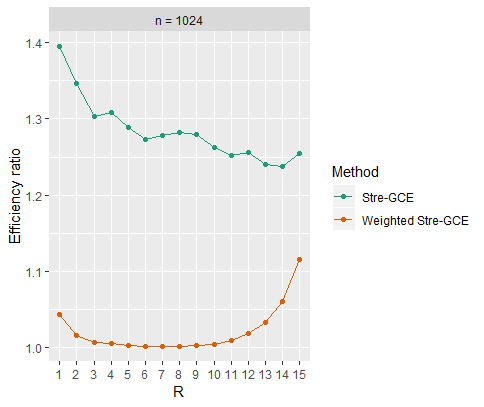} 
  \end{minipage} 
  \begin{minipage}[b]{0.48\linewidth}
    \centering
    \includegraphics[width=\linewidth]{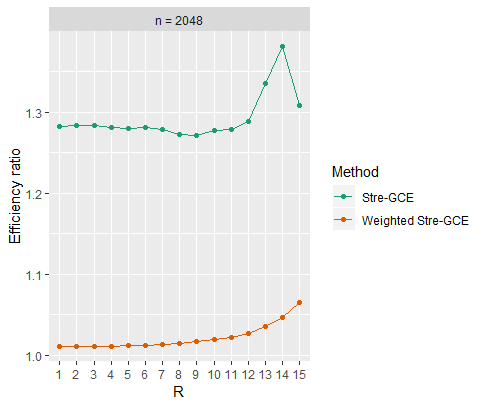} 
  \end{minipage}
  \begin{minipage}[b]{0.48\linewidth}
    \centering
    \includegraphics[width=\linewidth]{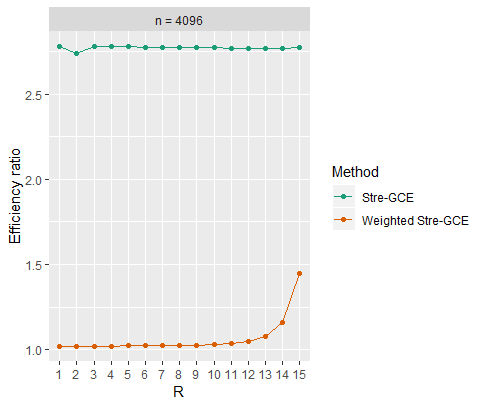} 
  \end{minipage} 
  \begin{centering}
  \caption{\label{fig: Stre-GCE performances} Performance of standard Stre-GCE
and Weighted Stre-GCE methods in terms of efficiencies ratios (\ref{eq: efficiency})
and (\ref{eq: efficiency weighted}).}
  \par\end{centering}
  \end{figure}

Based on this model, we performed a set of simulation to compare the
performance of Stre-GCE with those of Weighted Stre-GCE in terms of
the efficiency ratio 
\begin{equation}
\varrho_{\text{W}}:=\frac{\mathrm{RMSE}_{\text{Weighted Stre-GCE}}(\mathcal{D})}{\mathrm{RMSE}_{\text{GCE}}(\mathcal{D})}.\label{eq: efficiency weighted}
\end{equation}
For this purpose, we chose samples of sizes $n\in\{128,256,512,1024,2048,4096\}$ in order to carry out the performance analysis based on the procedure described below. Each dataset has been split into $16$ equal subsets. Then, for each $R\in\{1,\dots,15\}$, the first $R$ part was used as batch, while the remaining ones made up the streaming phase. Results of this extended set of simulations are reported in Table \ref{tab: rmse weighted Stre-GCE}, as well as graphically in Figure \ref{fig: Stre-GCE performances} where both methods are compared in terms of efficiency.

\begin{table}
\centering{}
\begin{subtable}[b]{0.3\textwidth}
\begin{tabular}{|c|c|c|} 
\hline
$R$  & $\varrho$ & $\varrho_{W}$ \\ \hline 
1  & 1,9284    & 1,4344        \\ \hline 
2  & 1,9494    & 1,4646        \\ \hline 
3  & 1,9976    & 1,5129        \\ \hline 
4  & 1,8993    & 1,4770        \\ \hline 
5  & 1,8980    & 1,4992        \\ \hline 
6  & 1,8561    & 1,4998        \\ \hline
7  & 1,8793    & 1,5394        \\ \hline 
8  & 1,8812    & 1,5699        \\ \hline 
9  & 1,8595    & 1,5916        \\ \hline 
10 & 1,8623    & 1,6308        \\ \hline 
11 & 1,8289    & 1,6476        \\ \hline 
12 & 1,7934    & 1,6625        \\ \hline 
13 & 1,7278    & 1,6426        \\ \hline 
14 & 1,8244    & 1,7564        \\ \hline 
15 & 1,7607    & 1,7333        \\ \hline 
\end{tabular}
\caption{$n=128$, $\mathrm{RMSE}=1,0407$}
\end{subtable}
\hfill
\begin{subtable}[b]{0.3\textwidth}
\begin{tabular}{|c|c|c|} 
\hline
$R$  & $\varrho$ & $\varrho_{W}$ \\ \hline
1  & 1,1757    & 1,1624        \\ \hline
2  & 1,1426    & 1,0998        \\ \hline
3  & 1,1367    & 1,0889        \\ \hline
4  & 1,1430    & 1,1040        \\ \hline
5  & 1,1341    & 1,0841        \\ \hline
6  & 1,1402    & 1,0933        \\ \hline
7  & 1,1445    & 1,1028        \\ \hline
8  & 1,1462    & 1,1082        \\ \hline
9  & 1,1434    & 1,1070        \\ \hline
10 & 1,1341    & 1,0962        \\ \hline
11 & 1,1325    & 1,0972        \\ \hline
12 & 1,1324    & 1,1015        \\ \hline
13 & 1,1331    & 1,1092        \\ \hline
14 & 1,1360    & 1,1099        \\ \hline
15 & 1,1167    & 1,0903        \\ \hline
\end{tabular}
\caption{$n=256$, $\mathrm{RMSE}=1,0062$}
\end{subtable}
\hfill
\begin{subtable}[b]{0.3\textwidth}
\begin{tabular}{|c|c|c|}  
\hline 
$R$  & $\varrho$ & $\varrho_{W}$ \\ \hline 
1  & 1,2574    & 1,0344        \\ \hline 
2  & 1,2264    & 1,0387        \\ \hline 
3  & 1,2212    & 1,0540        \\ \hline 
4  & 1,2231    & 1,0410        \\ \hline 
5  & 1,2234    & 1,0368        \\ \hline 
6  & 1,2255    & 1,0272        \\ \hline 
7  & 1,2255    & 1,0224        \\ \hline 
8  & 1,2259    & 1,0168        \\ \hline 
9  & 1,2282    & 1,0108        \\ \hline 
10 & 1,2279    & 1,0092        \\ \hline 
11 & 1,2281    & 1,0106        \\ \hline 
12 & 1,2297    & 1,0194        \\ \hline 
13 & 1,2302    & 1,0405        \\ \hline 
14 & 1,2321    & 1,0840        \\ \hline 
15 & 1,2355    & 1,1504        \\ \hline 
\end{tabular} 
\caption{$n=512$, $\mathrm{RMSE}=0,9485$}
\end{subtable}
\hfill
\begin{subtable}[b]{0.3\textwidth}
\begin{tabular}{|c|c|c|}
\hline
$R$  & $\varrho$ & $\varrho_{W}$ \\ \hline
1  & 1,3942    & 1,0431        \\ \hline
2  & 1,3462    & 1,0167        \\ \hline
3  & 1,3036    & 1,0066        \\ \hline
4  & 1,3089    & 1,0056        \\ \hline
5  & 1,2886    & 1,0032        \\ \hline
6  & 1,2730    & 1,0020        \\ \hline
7  & 1,2788    & 1,0017        \\ \hline
8  & 1,2816    & 1,0017        \\ \hline
9  & 1,2795    & 1,0025        \\ \hline
10 & 1,2628    & 1,0049        \\ \hline
11 & 1,2521    & 1,0098        \\ \hline
12 & 1,2557    & 1,0185        \\ \hline
13 & 1,2402    & 1,0337        \\ \hline
14 & 1,2373    & 1,0604        \\ \hline
15 & 1,2542    & 1,1163        \\ \hline
\end{tabular}
\caption{$n=1024$, $\mathrm{RMSE}=1,0306$}
\end{subtable}
\hfill
\begin{subtable}[b]{0.3\textwidth}
\begin{tabular}{|c|c|c|} 
\hline
$R$  & $\varrho$ & $\varrho_{W}$ \\ \hline
1  & 1,2827    & 1,0110        \\ \hline
2  & 1,2833    & 1,0110        \\ \hline
3  & 1,2833    & 1,0110        \\ \hline
4  & 1,2808    & 1,0112        \\ \hline
5  & 1,2792    & 1,0117        \\ \hline
6  & 1,2811    & 1,0126        \\ \hline
7  & 1,2788    & 1,0137        \\ \hline
8  & 1,2718    & 1,0150        \\ \hline
9  & 1,2715    & 1,0168        \\ \hline
10 & 1,2774    & 1,0192        \\ \hline
11 & 1,2780    & 1,0226        \\ \hline
12 & 1,2886    & 1,0275        \\ \hline
13 & 1,3347    & 1,0351        \\ \hline
14 & 1,3802    & 1,0471        \\ \hline
15 & 1,3084    & 1,0655        \\ \hline
\end{tabular}
\caption{$n=2048$, $\mathrm{RMSE}=0,9962$}
\end{subtable}
\hfill
\begin{subtable}[b]{0.3\textwidth}
\begin{tabular}{|c|c|c|} 
\hline
$R$  & $\varrho$ & $\varrho_{W}$ \\ \hline
1  & 2,7767    & 1,0186        \\ \hline
2  & 2,7373    & 1,0188        \\ \hline
3  & 2,7767    & 1,0197        \\ \hline
4  & 2,7766    & 1,0205        \\ \hline
5  & 2,7766    & 1,0218        \\ \hline
6  & 2,7765    & 1,0220        \\ \hline
7  & 2,7764    & 1,0225        \\ \hline
8  & 2,7729    & 1,0239        \\ \hline
9  & 2,7711    & 1,0253        \\ \hline
10 & 2,7715    & 1,0280        \\ \hline
11 & 2,7690    & 1,0334        \\ \hline
12 & 2,7690    & 1,0454        \\ \hline
13 & 2,7683    & 1,0753        \\ \hline
14 & 2,7661    & 1,1585        \\ \hline
15 & 2,7731    & 1,4466        \\ \hline
\end{tabular}
\caption{$n=4096$, $\mathrm{RMSE}=0,9992$}
\end{subtable}
\vspace*{0.2cm}
\caption{\label{tab: rmse weighted Stre-GCE} Comparison of efficiencies associated
with the standard Stre-GCE and the weighted Stre-GCE methods. The
efficiency ratios (\ref{eq: efficiency}) for standard Stre-GCE and
(\ref{eq: efficiency weighted}) for Weighted Stre-GCE are shown at
datasets of sizes $n\in\{128,256,512,1024,2048,4096\}$. Each dataset
was divided in $16$ parts with equal cardinality: for each $R\in\{1,\dots,15\}$,
an associated simulation test was run using the first $r$ parts as
batch and the remaining $16-R$ parts for the streaming phase.}
\end{table}

The same analysis described above has been carried out on datasets of sizes $n\in\{480,960,1920,3840\}$ already considered in Table \ref{tab: rmse at sample sizes 60, 120, 240, 480, 960, 1920, 3840}, in order to explore the change in the performance when moving from standard Stre-GCE to Weighted Stre-GCE. The results are presented in Figure \ref{fig: Stre-GCE performances, 2}.  

\begin{figure}[ht]   
  \begin{minipage}[b]{0.48\linewidth}
    \centering
    \includegraphics[width=\linewidth]{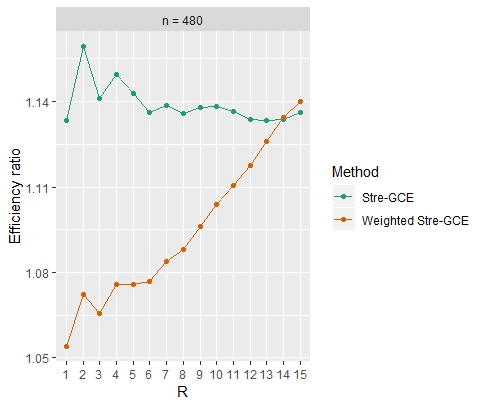} 
  \end{minipage} 
  \begin{minipage}[b]{0.48\linewidth}
    \centering
    \includegraphics[width=\linewidth]{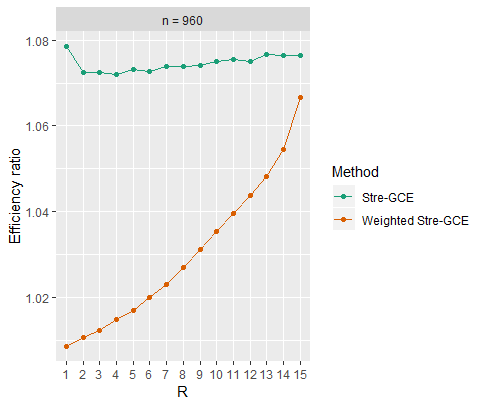} 
  \end{minipage} 
  \begin{minipage}[b]{0.48\linewidth}
    \centering
    \includegraphics[width=\linewidth]{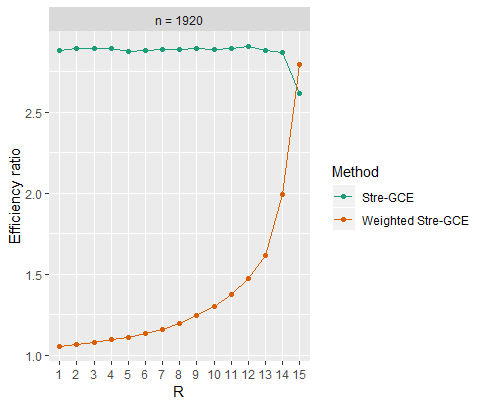} 
  \end{minipage}
  \begin{minipage}[b]{0.48\linewidth}
    \centering
    \includegraphics[width=\linewidth]{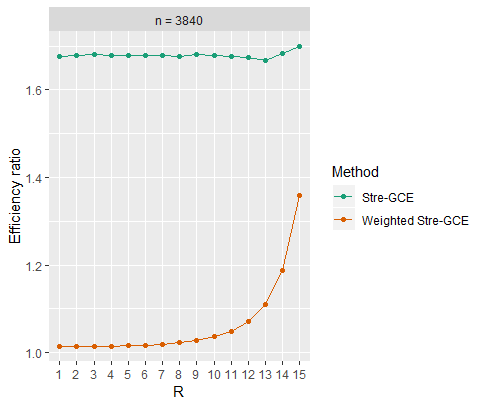} 
  \end{minipage} 
  \begin{centering}
  \caption{\label{fig: Stre-GCE performances, 2} Performance of standard Stre-GCE
and Weighted Stre-GCE methods in terms of efficiencies ratios (\ref{eq: efficiency})
and (\ref{eq: efficiency weighted}).}
  \par\end{centering}
  \end{figure}

\section{Discussion }

\label{sec: Discussion }

We summarise some of the features that can be extracted from the previous
observations and the set of simulations.

As already remarked, the order of observations for updates of cross
entropy is irrelevant when the update involves all the data at the
same time (standard GCE), whereas Stre-GCE gives relevance to the temporal structure, i.e.
the timeflow represented by the data stream. This additional
structure allows to distinguish between two sources of entropy and
is linked to the relaxation of consistency constraints (as remarked
in Section \ref{sec: Characteristics and range of application}).
On the other hand, results are not compromised by multicollinearity
in contrast to other recursive approaches, as can be expected from
the broad scope of applicability of entropy-based methods. 

The timeflow of data also raises other concerns: for instance, standardization assumes the
knowledge of the full dataset, therefore it can be applied only to
the available data in a streaming context. Alternatively, the
performance of the Stre-GCE method on the full standardized dataset
can only be used \textit{a posteriori}, which is the approach we adopted to get the results shown in Table \ref{tab: rmse at sample sizes 60, 120, 240, 480, 960, 1920, 3840}.
It is worth remarking that the simulation we carried out with standardized
variables shows a reduced dependence on the batch size: the differences
among RMSEs at different batch sizes are often smaller than the last
significant digit shown in Table \ref{tab: rmse at sample sizes 60, 120, 240, 480, 960, 1920, 3840}. 

We have considered a ``hybrid'' approach through blocks, in order
to interpolate between Stre-GCE ($g=1$) and standard GME ($g=n$).
Indeed, the relationship between GME and Stre-GCE can be seen as a
specific application of the MEP methods at different \emph{time scales}.
In physical terms, the assumption (\ref{eq: consistency constraints GME})
of variables being replaced by their associated first moments, is
valid under certain conditions, often including the characteristic
time scales of the model (e.g., thermalization time). Hence, blocks
at $g\notin\{1,n\}$ represent a new approach to deal with the notion
of \emph{metastability} also discussed in a different framework in
\cite{Angelelli2017,AK2016b}. We stress that, in applications, streaming
blocks are not only an additional tool to refine the algorithm, but
can characterize specific models where data clusters add in a sequential
way.

We note that Stre-GCE is more sensitive to the dataset than standard
GCE with uniform prior: this sensitivity does not only rely on the
data, but also on their ordering within the stream. Different
factors may result in the distortion of the estimates, ranging from the
consolidation of fluctuations during the streaming phase, to excessive
importance linked to streaming data. In order to avoid
data overrating during the streaming phase, yet preserving the temporal structure,
an alternative streaming algorithm has been proposed (i.e. Weighted
Stre-GCE, see Section \ref{sec: Weighted Stre-GCE}). Specifically, the
balancing of the relaxation of consistency constraints with the inclusion of new observation
has led to the introduction of weights (\ref{eq: weighted update step, estimates}). These weights are \emph{natural} in that they can be described as a recursive relationship linking the objective function at the $i^{\text{th}}$ step to that at the $(i+1)^{\text{th}}$ step independently
on the dataset. The quality of the estimates is consistently improved
when compared with the simple Stre-GCE method, as it can be seen in
the comparison Table \ref{tab: rmse weighted Stre-GCE}. In particular,
the deviations from a monotone behaviour of the RMSE as the data dimension
increases are mitigated in the Weighted Stre-GCE approach and results are
consistent with those for standard GCE: in this regard, weights act
as a ``smoothing'' factor.

When the batch dimension is large, the standard Stre-GCE and Weighted Stre-GCE methods tend to produce similar results. Furthermore, it is noted that a short streaming phase in the Weighted Stre-GCE
approach gives more relevance to the individual data during the stream,
so they may affect the performance of the method. In fact, for each fixed dataset, the larger the streaming, the smaller the batch and the RMSE associated with Weighted Stre-GCE tends to
increase along with the size of the batch (also see Figures \ref{fig: Stre-GCE performances} and \ref{fig: Stre-GCE performances, 2}).
New data can suppress fluctuations, resulting into a tendency to perform
better on larger datasets, area in which both GME and GCE struggle
the most from a computational point of view. 

One way to improve the results and avoid the decline in performance is to enter a control step in the Stre-GCE algorithm, that is, the choice of whether or not to update the distribution. If we fix a performance indicator $\Delta(\boldsymbol{\hat{\beta}})$ for the estimate $\boldsymbol{\hat{\beta}}$, at the $(i+1)^{\text{th}}$ iteration step we can choose between either the last available estimate $\boldsymbol{\hat{\beta}}_{i}$ or the new potential estimate $\boldsymbol{\hat{\beta}}_{i+1}^{(\text{new)}}$ obtained from the update rules (\ref{eq: update step, estimates}) or (\ref{eq: weighted update step, estimates}). The choice is based on the following criterion 
\begin{equation}
\Delta(\boldsymbol{\hat{\beta}}_{i+1})=\min\left\{ \Delta(\boldsymbol{\hat{\beta}}_{i}),\Delta(\boldsymbol{\hat{\beta}}_{i+1}^{(\text{new)}})\right\}.
\label{eq: update criteria performance indicator}
\end{equation}
which only relies on the data available at the $(i+1)^{\text{th}}$ step.
However, in this work we focused on the RMSE as performance indicator for \textit{a posteriori} evaluation, always involving the whole dataset. Another quantity of interest is represented by
\emph{entropy production}, which is related to nonequilibrium processes.
In particular, we can consider the evolution for the model itself,
which may vary during the data stream. For instance, the parameters
$\boldsymbol{\beta}$ may change, or nonlinearities could appear outside
a given time scale. In these cases, one cannot rely on the approximate
estimate provided by GCE applied on the batch only, as new data may
contain relevant information on the parameters and their evolution.
We will consider these approaches in a separate paper.

Finally, we remark that Stre-GCE assigns asymmetric roles to signal
and error in an intrinsic way, namely, without the introduction of
any additional parameter to be estimated. The Weighted Stre-GCE presented
in this work is in line with such a characteristic: the weights $w_{\beta}^{(i)},w_{\varepsilon}^{(i)}$
in (\ref{eq: weighted KL objective function}) are derived solely
from the definition of cardinality of sets and from the different
contributions attributed to the signal and the error during the updates.
A similar approach has been formalized also in the context of statistical
physics using algebraic methods \cite{Angelelli2017}.

\section{Conclusion and future perspectives}

\label{sec: Conclusion and future perspectives}

This work aimed at introducing the Stre-GCE method to combine an adaptive
approach with generalized entropy methods. The qualitative discussion
and the quantitative analysis that have been performed suggest several
extensions and applications for future investigations, some of which
have already been mentioned.

Following the observations in Section \ref{sec: Discussion }, the algorithm
can be improved choosing between the update or the preservation of
the actual distribution $\mathbf{P}^{\beta}$ at each step. In this
regards, either the Root-Mean-Square Error or the entropy production
represent criteria that might drive the decision process. Additionally,
we remark that the comparison between Stre-GCE and standard GCE can
be used \textit{a posteriori} to provide a finer description of blocks
of data in the flow of time in terms of the quality of the estimates.
%In fact, the possibility to quantify the amount of entropy production for new (blocks of) data may serve as a criterion to group similar observations and to identify redundant units or outliers.

We can consider more general forms of recursive weights between the
contributions of signal and error, that is a sequence $\gamma_{i}\in[0,1]$,
$i\in\{1,\dots,n\}$, a deformation of the objective function (\ref{eq: KL objective function})
defined by $\gamma_{i+1}\cdot\mathrm{D_{KL}}(\mathbf{P}^{\beta}||\mathbf{\hat{P}}_{i}^{\beta})+(1-\gamma_{i+1})\cdot\mathrm{D_{KL}}(\mathbf{p}_{i+1}^{\varepsilon}||\mathbf{u}_{i}^{\varepsilon})$
and a given rule $\gamma_{i+1}=f_{i}(\gamma_{i})$. Depending on the
choices of $f_{i}$, one can describe different balancing modes between
prior information and new data. With respect to this, the dependence
of the performance of the weighted algorithm in terms of initial weights
and update rule should be investigated too. 

The streaming approach is best suited to models that show a strong
temporal component. Therefore, it seems natural to adapt the Stre-GCE
algorithm and its generalizations to time-varying models, e.g. considering
a time-varying parameter $\beta$ in the relation (\ref{eq: matrix consistency constraint}).
This framework also suggests to further explore the complementary aspects of both the batch and the streaming phases as the information extraction from data goes on: in fact, they may alternate during the time window under consideration. This situation may be analyzed using a well-known approach to explore
time asymmetry, namely, non-commutativity of two \emph{learning modes}, that is, the batch learning mode and the streaming learning mode.

Finally, applications to real systems will be considered in next investigations.
The learning aspects, also including recursive weights, and the entropy
production could be used in both dynamical frameworks such as those
observed in the psychological sciences (e.g. the psychotherapeutic
process that aims at change and is expected to evolve over time) and
in the context of reasoning under uncertainty of knowledge-based social
networks: both of them provide natural settings where the sequential
data acquisition, the evolution over time, the combination of prior
and new information, as well as their quantification play a fundamental
role.

 %{\footnotesize
\bibliographystyle{elsarticle-num}
\bibliography{Streaming_Generalised_Cross_Entropy_ARXIV}

\begin{thebibliography}{10}
\expandafter\ifx\csname url\endcsname\relax
  \def\url#1{\texttt{#1}}\fi
\expandafter\ifx\csname urlprefix\endcsname\relax\def\urlprefix{URL }\fi
\expandafter\ifx\csname href\endcsname\relax
  \def\href#1#2{#2} \def\path#1{#1}\fi

\bibitem{Solomonoff1964}
R.~J. Solomonoff, {A} formal theory of inductive inference, {P}arts i, ii,
  Inform. control 7~(1, 2) (1964) 1--22, 224--254.
\newblock \href {https://doi.org/10.1016/s0019-9958(64)90131-7}
  {\path{doi:10.1016/s0019-9958(64)90131-7}}.

\bibitem{Holzinger2014}
A.~Holzinger, M.~H\"{o}rtenhuber, C.~Mayer, M.~Bachler, S.~Wassertheurer, A.~J.
  Pinho, D.~Koslicki, {O}n {E}ntropy-{B}ased {D}ata {M}ining, in: {I}nteractive
  {K}nowledge {D}iscovery and {D}ata {M}ining in {B}iomedical {I}nformatics,
  Springer Nature, 2014, pp. 209--226.
\newblock \href {https://doi.org/10.1007/978-3-662-43968-5_12}
  {\path{doi:10.1007/978-3-662-43968-5_12}}.

\bibitem{Berger1996}
A.~L. Berger, V.~J. Della~Pietra, S.~Della~Pietra, {A} maximum entropy approach
  to natural language processing, Comput. Linguist. 22~(1) (1996) 39--71.

\bibitem{Gladyshev1997}
G.~Gladyshev, {T}hermodynamic {T}heory of the {E}volution of {L}iving {B}eings,
  Nova Science Pub. Inc., 1997.

\bibitem{Golan1996}
A.~Golan, G.~Judge, D.~Miller, {M}aximum entropy econometrics: {R}obust
  estimation with limited data, Chichester, UK: John Wiley \& Sons, 1996.

\bibitem{Golan2007}
A.~Golan, {I}nformation and {E}ntropy {E}conometrics -- {A} {R}eview and
  {S}ynthesis, Found. Trends Econom. 2~(1-2) (2007) 1--145.
\newblock \href {https://doi.org/10.1561/0800000004}
  {\path{doi:10.1561/0800000004}}.

\bibitem{Golan2018}
A.~Golan, {F}oundations of {I}nfo-{M}etrics: {M}odeling, {I}nference, and
  {I}mperfect {I}nformation, New York, NY: Oxford University Press, 2018.
\newblock \href {https://doi.org/10.1093/oso/9780199349524.001.0001}
  {\path{doi:10.1093/oso/9780199349524.001.0001}}.

\bibitem{Ciavolino2009a}
E.~Ciavolino, J.~J. Dahlgaard, {S}imultaneous equation model based on the
  generalized maximum entropy for studying the effect of management factors on
  enterprise performance, J. Appl. Stat. 36~(7) (2009) 801--815.
\newblock \href {https://doi.org/10.1080/02664760802510026}
  {\path{doi:10.1080/02664760802510026}}.

\bibitem{Ciavolino2009b}
E.~Ciavolino, A.~D. Al-Nasser, {C}omparing generalised maximum entropy and
  partial least squares methods for structural equation models, J. Nonparametr.
  Stat. 21~(8) (2009) 1017--1036.
\newblock \href {https://doi.org/10.1080/10485250903009037}
  {\path{doi:10.1080/10485250903009037}}.

\bibitem{Ciavolino2014a}
E.~Ciavolino, A.~Calcagn\`{i}, {G}eneralized cross entropy method for analysing
  the {SERVQUAL} model, J. Appl. Stat. 42~(3) (2014) 520--534.
\newblock \href {https://doi.org/10.1080/02664763.2014.963526}
  {\path{doi:10.1080/02664763.2014.963526}}.

\bibitem{Ciavolino2014b}
E.~Ciavolino, M.~Carpita, {T}he {GME} estimator for the regression model with a
  composite indicator as explanatory variable, Qual. Quant. 49~(3) (2014)
  955--965.
\newblock \href {https://doi.org/10.1007/s11135-014-0061-4}
  {\path{doi:10.1007/s11135-014-0061-4}}.

\bibitem{Amusa2019}
L.~Amusa, T.~Zewotir, D.~North, Examination of entropy balancing technique for
  estimating some standard measures of treatment effects: a simulation study,
  Electron. J. Appl. Stat. Anal. 12~(2) (2019) 491--507.
\newblock \href {https://doi.org/10.1285/i20705948v12n2p491}
  {\path{doi:10.1285/i20705948v12n2p491}}.

\bibitem{Bernardini2011}
R.~Bernardini~Papalia, E.~Ciavolino, {GME} estimation of spatial structural
  equations models, J. Classif. 28~(1) (2011) 126--141.
\newblock \href {https://doi.org/10.1007/s00357-011-9073-0}
  {\path{doi:10.1007/s00357-011-9073-0}}.

\bibitem{Ciavolino2014c}
E.~Ciavolino, M.~Carpita, A.~D. Al-Nasser, {M}odelling the quality of work in
  the italian social co-operatives combining {NPCA-RSM} and {SEM-GME}
  approaches, J. Appl. Stat. 42~(1) (2014) 161--179.
\newblock \href {https://doi.org/10.1080/02664763.2014.938226}
  {\path{doi:10.1080/02664763.2014.938226}}.

\bibitem{Ciavolino2016}
E.~Ciavolino, A.~Calcagn\`{i}, {A} generalized maximum entropy ({GME})
  estimation approach to fuzzy regression model, Appl. Soft Comput. 38 (2016)
  51--63.
\newblock \href {https://doi.org/10.1016/j.asoc.2015.08.061}
  {\path{doi:10.1016/j.asoc.2015.08.061}}.

\bibitem{Simon2006}
D.~Simon, {O}ptimal {S}tate {E}stimation, Hoboken, NJ: John Wiley {\&} Sons,
  2006.
\newblock \href {https://doi.org/10.1002/0470045345}
  {\path{doi:10.1002/0470045345}}.

\bibitem{Widrow1988}
B.~Widrow, R.~Winter, {N}eural nets for adaptive filtering and adaptive pattern
  recognition, Computer 21~(3) (1988) 25--39.
\newblock \href {https://doi.org/10.1109/2.29} {\path{doi:10.1109/2.29}}.

\bibitem{Xu2002}
X.~Xu, H.~He, D.~Hu, {E}fficient {R}einforcement {L}earning {U}sing {R}ecursive
  {L}east-{S}quares {M}ethods, J. Artif. Intell. Res. 16 (2002) 259--292.
\newblock \href {https://doi.org/10.1613/jair.946}
  {\path{doi:10.1613/jair.946}}.

\bibitem{Zanetti2012}
R.~Zanetti, {R}ecursive {U}pdate {F}iltering for {N}onlinear {E}stimation, IEEE
  T. Automat. Contr. 57~(6) (2012) 1481--1490.
\newblock \href {https://doi.org/10.1109/tac.2011.2178334}
  {\path{doi:10.1109/tac.2011.2178334}}.

\bibitem{Daum2005}
F.~Daum, {N}onlinear filters: beyond the {K}alman filter, IEEE Aero. El. Sys.
  Mag. 20~(8) (2005) 57--69.
\newblock \href {https://doi.org/10.1109/maes.2005.1499276}
  {\path{doi:10.1109/maes.2005.1499276}}.

\bibitem{Bagya2018}
H.~Bagya~Lakshmi, M.~Gallo, R.~M. Srinivasan, Comparison of regression models
  under multi-collinearity, Electron. J. Appl. Stat. Anal. 11~(1) (2018)
  340--368.
\newblock \href {https://doi.org/10.1285/i20705948v11n1p340}
  {\path{doi:10.1285/i20705948v11n1p340}}.

\bibitem{LL1980}
L.~D. Landau, E.~M. Lifschitz, {S}tatistical {P}hysics, Vol.~5 of Course of
  Theoretical Physics, Oxford, UK: Butterworth-Heinemann, 1980.

\bibitem{Feynman1982}
R.~P. Feynman, {S}tatistical {M}echanics: {A} {S}et {O}f {L}ectures, Advanced
  Books Classics (revised edition), Boulder, CO: Westview Press, 1998.

\bibitem{Cover2006}
T.~M. Cover, J.~A. Thomas, {E}lements of {I}nformation {T}heory, Hoboken, NJ:
  John Wiley \& Sons, 2006.
\newblock \href {https://doi.org/10.1002/047174882x}
  {\path{doi:10.1002/047174882x}}.

\bibitem{Khinchin1957}
A.~I. Khinchin, {M}athematical {F}oundations of {I}nformation {T}heory, Dover,
  1957.

\bibitem{Dewar2009}
R.~Dewar, {M}aximum {E}ntropy {P}roduction as an {I}nference {A}lgorithm that
  {T}ranslates {P}hysical {A}ssumptions into {M}acroscopic {P}redictions:
  {D}on't {S}hoot the {M}essenger, Entropy 11~(4) (2009) 931--944.
\newblock \href {https://doi.org/10.3390/e11040931}
  {\path{doi:10.3390/e11040931}}.

\bibitem{Jaynes2003}
E.~T. Jaynes, {P}robability {T}heory - The {L}ogic of {S}cience, Cambridge, UK:
  Cambridge University Press, 2003.
\newblock \href {https://doi.org/10.1017/CBO9780511790423}
  {\path{doi:10.1017/CBO9780511790423}}.

\bibitem{Golan1998}
A.~Golan, ``{M}aximum entropy, likelihood and uncertainty'', in: G.~Erickson,
  J.~T. Rychert, C.~R. Smith (Eds.), {M}aximum {E}ntropy and {B}ayesian
  {M}ethods. Boise, Idaho, USA, 1997: Proceedings of the 17th International
  Workshop on Maximum Entropy and Bayesian Methods of Statistical Analysis,
  Vol.~98 of Fundamental Theories of Physics, Netherlands: Springer, 1998.
\newblock \href {https://doi.org/10.1007/978-94-011-5028-6}
  {\path{doi:10.1007/978-94-011-5028-6}}.

\bibitem{Bertsekas1975}
D.~P. Bertsekas, {C}ombined primal-dual and penalty methods for constrained
  minimization, SIAM J. Control 13~(3) (1975) 521--544.
\newblock \href {https://doi.org/10.1137/0313030} {\path{doi:10.1137/0313030}}.

\bibitem{Wu2009}
X.~Wu, {A} {W}eighted {G}eneralized {M}aximum {E}ntropy {E}stimator with a
  {D}ata-driven {W}eight, Entropy 11~(4) (2009) 917--930.
\newblock \href {https://doi.org/10.3390/e11040917}
  {\path{doi:10.3390/e11040917}}.

\bibitem{Jarzynski1997}
C.~Jarzynski, {N}onequilibrium {E}quality for {F}ree {E}nergy {D}ifferences,
  Phys. Rev. Lett. 78~(14) (1997) 2690--2693.
\newblock \href {https://doi.org/10.1103/physrevlett.78.2690}
  {\path{doi:10.1103/physrevlett.78.2690}}.

\bibitem{Crooks1999}
G.~E. Crooks, {E}ntropy production fluctuation theorem and the nonequilibrium
  work relation for free energy differences, Phys. Rev. E 60~(3) (1999)
  2721--2726.
\newblock \href {https://doi.org/10.1103/physreve.60.27}
  {\path{doi:10.1103/physreve.60.27}}.

\bibitem{Pukelsheim1994}
F.~Pukelsheim, {T}he {T}hree {S}igma {R}ule, Am. Stat. 48~(2) (1994) 88--91.
\newblock \href {https://doi.org/10.2307/2684253} {\path{doi:10.2307/2684253}}.

\bibitem{Angelelli2017}
M.~Angelelli, {T}ropical limit and micro-macro correspondence in statistical
  physics, J. Phys. A: Math. Theor. 50 (2017) 415202.
\newblock \href {https://doi.org/10.1088/1751-8121/aa863b}
  {\path{doi:10.1088/1751-8121/aa863b}}.

\bibitem{AK2016b}
M.~Angelelli, B.~Konopelchenko, {Z}eros and amoebas of partition functions,
  Rev. Math. Phys. 30~(09) (2018) 1850015.
\newblock \href {https://doi.org/10.1142/s0129055x18500150}
  {\path{doi:10.1142/s0129055x18500150}}.

\end{thebibliography}
 %}
\end{document}